\documentclass[useAMS,usenatbib]{mn2e} 
\ifx\pdfoutput\undefined
\usepackage{graphicx}
\else
\usepackage[pdftex]{graphicx}
\usepackage{epstopdf}
\fi

\usepackage{psfrag}
\usepackage{color}

\def\solmas{$\mathrm{M_\odot}$~}
\def\solmasp{$\mathrm{M_\odot}$}
\def\tcr{$t_{\mathrm{cr}}$}
\def\sims{$\sim$}
\def\Vff{$f$}
\def\Vffpic{$\mathbf{f}$} 
\def\tcrpic{$\mathbf{t_{\mathrm{cr}}}$} 
\def\Machpic{$\mathbf{\mathcal{M}}$}
\def\Mach{$\mathcal{M}$}

 \title[Clumpy shocks and the clump mass function]{Clumpy shocks and the clump mass function}
 \author[Clark \& Bonnell]
 {Paul C. Clark \thanks{E-mail: pcc@st-andrews.ac.uk} \& Ian A. Bonnell\\ School of Physics \&
Astronomy, University of St Andrews, North Haugh, St Andrews, Fife, KY16 9SS. \\ }

\date{\today}

\begin{document}
\maketitle

%
% Abstract....
%

\begin{abstract}
One possible mechanism for the formation of molecular clouds is large scale colliding
flows.  In this paper, we examine whether clumpy, colliding, flows could be responsible
for the clump mass functions that have been observed in several regions of embedded
star formation, which have been shown to be described
by a Salpeter type slope. The flows presented here, which comprise a population of 
initially identical
clumps, are modelled using smoothed particle hydrodynamics (SPH) and calculations
are performed with and without the inclusion of self-gravity. When the shock region
is at its densest, we find that the clump mass spectrum is always well modelled by a 
Salpeter type slope. This is
true regardless of whether the self-gravity is included in the simulations or not, and for
our choice of filling
factors for the clumpy flows (10, 20 \& 40\%), and Mach number (5, 10 \& 20). In the
non-self-gravitating simulations, this slope is retained at lower Mach numbers 
(\Mach = 5 \& 10) as the simulations progress past the densest phase. In the simulations 
which include self-gravity, we find that low Mach number runs yield a flatter mass function
after the densest phase. This is simply a result of increased coagulation due to gravitational 
collapse of the flows. In the high Mach number runs (\Mach = 20) the Salpeter slope is 
always lost. The self gravitating calculations also show that the sub-group of gravitationally 
bound  clumps in which star formation occurs, always contain the most massive clumps in the
population. Typically these clumps have a mass of order of the Jeans mass of the initial clumps.
The mass function of these bound star forming clumps is not at all similar to 
the Salpeter type mass function observed for stars in the field. We conclude that the clump 
mass function may not only have nothing to do with gravity, but also nothing to do with 
the star formation process and the resulting mass distribution of stars. This raises doubt over
the claims that the clump mass function is the origin of the stellar IMF, for regions such as 
$\rho$ Oph, Serpens and the Orion B cloud.
\end{abstract}

%
% Key words
%


%
% Background and motivation
%

\section{Introduction}

The mass spectrum of stars, clumps or clouds, is normally described by some sort of
power law, which has the form $dN \propto m^{-\gamma} dm$, where $N$ is the
number of objects, and $m$ their mass.
Recent sub-millimetre observations of clumps and cores in the star forming regions
of $\rho$ Ophiuchus, Serpens, and Orion B 
have shown that the mass distribution of these gas 
density enhancements is similar to the mass distribution of young stars 
\citep{Motteetal1998, TestiSargent1998, Johnstoneetal2000, Johnstoneetal2001}. 
More recent observations using the extinction mapping technique \citep{Alvesetal2001} 
have confirmed this result in a number of star forming regions \citep{Ladaetal2006}.
These observations have been presented as 
evidence that the fragmentation process controls the mass of stars. In this picture, a 
bound region of smooth gas breaks up into a series of fragments (the clumps or cores, 
depending on your terminology) as the gravitational instability sets in. These fragments 
then collapse to form individual stars.

There is however the suggestion that most clumps in this region appear to be stable
against gravitational collapse, at least in $\rho$ Oph and the Orion B cloud \citep{Johnstoneetal2000, Johnstoneetal2001}.  
We therefore do not know if the clumps that make up the mass distribution are 
bound objects, or if they are just transient features that will never collapse to form stars. 
It is possible that most of the density structures observed in these regions have little to 
do with the star formation process.

Such transient density features are commonly found in simulations of turbulent gas, 
leading to the suggestion that the clump mass spectra, and thus potentially the stellar
IMF, is controlled by shock density enhancements in turbulent gas \citep{PN2002}.
Numerical studies of driven and freely decaying turbulence show that a clump mass 
spectrum similar to the the IMF can be formed via the shocks arising from the turbulent 
flows (e.g. \citealt{Klessen2001, Ballesterosetal2005}).
 It is generally found that most of these clumps are unbound.
\citet{ClarkBonnell2005} have also argued that the clumpy structure, and thus the turbulent
motions themselves, are essentially irrelevant to the star formation process, with the mean
Jeans mass in the cloud controlling the star formation.

It is still unclear however whether the `turbulence' modelled in the above star 
formation simulations is an accurate description of the velocities inside molecular clouds.
For example, molecular clouds, and the giant molecular clouds (GMCs) where most
star formation occurs, may not be coherent structures, but rather an ensemble of smaller
clouds which have been accumulated by some external mechanism \citep{Bonnelletal2005}.
At the point of a cloud's 
`formation', the smaller clouds out of which it comprises will probably have some random 
motions. In this model, the internal structure and velocity of a parent molecular cloud is then
dictated by the state of the constituent clouds. Regions of high density, such as the regions
where stars can form, will therefore be the result of clump-clump collisions, or of larger 
collisions between coherent flows of clumpy material.

The collisions between the internal structures of these molecular clouds (the original 
smaller clouds) have been suggested as a possible trigger for star formation
\citep{Smith1980}.
There are extensive studies of cloud-cloud collisions in the literature, involving numerical
calculations of collisions between pairs of clouds. These have looked at different Mach 
numbers, collisions between clouds of different sizes, gas cooling during the shocks, 
and the appearance of young embedded protostars
\citep{Chapmanetal1992, Bhattaletal1998}. The results of these simulations suggest
that is should be possible to form a wide range of binary separations and component
masses via random collisions between clumps of varying size/mass etc.
Only one study to date however, \citet{Gittinsetal2003}, has modelled an 
ensemble of such `cloudlets' using a fluid code. The purpose of that study was to 
examine whether the coagulation process can lead to a stellar type clump IMF, assuming
a system of initially identical cloudlets, with random motions. They found that very few 
mergers occurred in their simulations (due to the assumed small filling factor), 
resulting in a final mass spectrum of cloudlets that has a slope somewhat steeper 
than that of the Salpeter IMF.

Cloud-cloud collisions have also been investigated analytically, via the coagulation 
equation (e.g. \citealt{FieldSaslaw1965}; \citealt{FieldHutchins1968}; \citealt{Kwan1979}; \citealt{ScovilleHersh1979}; \citealt{SilkTakahashi1979}).
Assuming various combinations of cloud cross sections
and velocities, these studies concluded that the steady state mass distribution of the
clouds would have $\gamma \sim 1.5$. \citet{Penstonetal1969} and
\citet{Handburyetal1977}, using Monte-Carlo simulations, found $\gamma$ in the range
$1.5$ to $1.9$ respectively. 
However all these results assume that contact between two clouds
will result in a merger. Using 3D hydrodynamic simulations, \citet{Hausman1981}
demonstrated that not all collisions result in mergers, with high Mach numbers causing
the clouds to be shredded. \citet{Hausman1982} used these numerical collision 
calculations to include a prescription of the shredding/merging process into 
Monte-Carlo simulations. Again, the results showed that the mass spectrum should have 
$\gamma$ in the range $1.5$ to $2$. \citet{Elmegreen1989} included the effects of self-gravity 
into the coagulation equation, and how it can increase the chances of cloud-cloud collisions. 
His results showed that the inclusion of `gravitational focusing' of the collisions can result
in a steady-state cloud mass spectrum with $\gamma = 2$, but the calculations didn't 
include a prescription for possible shredding. In fact, by parameterising the various 
properties of the initial cloudlet population, one can get a wide range of final mass function
slopes \citep{SilkTakahashi1979}.

The results of both the analytic and the numerical studies have shown that a power law
type mass spectrum can be achieved from clump-clump (or cloud-cloud) collisions
between objects moving with random motions. It has been suggested however that
star forming regions are formed from large scale flows \citep{Semadenietal1995, 
Pringleetal2001}. If this is true, it is unlikely
that these flows would be of a uniform density, but instead be clumpy. In this scenario,
the chances of collisions are greatly increased. The resulting mass spectrum is also
unlikely to be steady state, as is found for the random motion studies, since the flows
have finite extent.

In this paper, we wish to examine whether the shock compressed 
regions created by supersonic, and clumpy,  flows can produce a clump/core mass 
function that resembles the stellar IMF. 
We wish also to examine the importance of gravity in building the clump/core population. 
We do not attempt here to fully explore the potentially vast parameter space associated
with clumpy flows but instead consider the simplest case of initially identical, uniform 
density, clumps. We thus only try to test the concept here, rather than build up a detailed
picture of cloud formation.
Both self-gravitating and non self-gravitating 
simulations are presented. In section \ref{calculations} we present the details of the 
simulations, including a discussion of the SPH code and the method of clump-finding that 
we have used. We go on to discuss the results of the non self-gravitating calculations in 
Section \ref{nogravsection} and the results of the simulations with self-gravity in Section 
\ref{gravsection}. There is a discussion of these results in relation to the $\rho$ Oph star 
forming region in Section \ref{discussion} and we conclude with a brief summary of our 
main findings in Section \ref{summary}.

\begin{table}
\caption{ \label{initialdetails}
Details of the nine simulations with no self-gravity are presented below.
The volume filling factor of the flow, \Vff, if defined as the percentage of the flow
volume that is occupied by gas. M$_{flow}$ is the total mass in the flow, and 
M$_{cloudlet}$ is the mass of the initial template clumps that are used to populate
the flow's volume, and r$_{cloudlet}$ is the radius of these clumps.
The calculations which include self-gravity have identical initial
conditions to those of simulations 7, 8 and 9.}
\begin{center}
\begin{tabular}{c|c|c|c|c|c}
\hline
Sim No. 	& \Vff 	& \Mach 	& M$_{flow}$ 	& M$_{cloudlet}$ 	& r$_{cloudlet}$	 \\
                   &(percent)& 	         & (\solmas)	& (\solmas) 		& (pc)		 \\
\hline \hline
1		& 10		& 5		& 1104		& 4.4				&  0.37	\\
2		& 10		& 10		& 1104		& 4.4				&  0.37	\\
3		& 10		& 20		& 1104		& 4.4				&  0.37	\\
4		& 20		& 5		& 1390		& 5.6				&  0.46	\\
5		& 20		& 10		& 1390		& 5.6				&  0.46	\\
6		& 20		& 20		& 1390		& 5.6				&  0.46	\\
7		& 40		& 5		& 1752		& 7.0				&  0.58	\\
8		& 40		& 10		& 1752		& 7.0				&  0.58	\\
9		& 40		& 20		& 1752		& 7.0				&  0.58	\\
\hline
\end{tabular}
\end{center}
\label{default}
\end{table}

%
% Figures.....
%

\begin{figure*}
\normalsize
{\bf{
\centerline{ 	\psfrag{time}{\textcolor{white}{t = 0.}}
			\psfrag{fill}{\textcolor{white}{\Vffpic ~= 10\%}}
		\includegraphics[width=2.1in,height=2.1in]{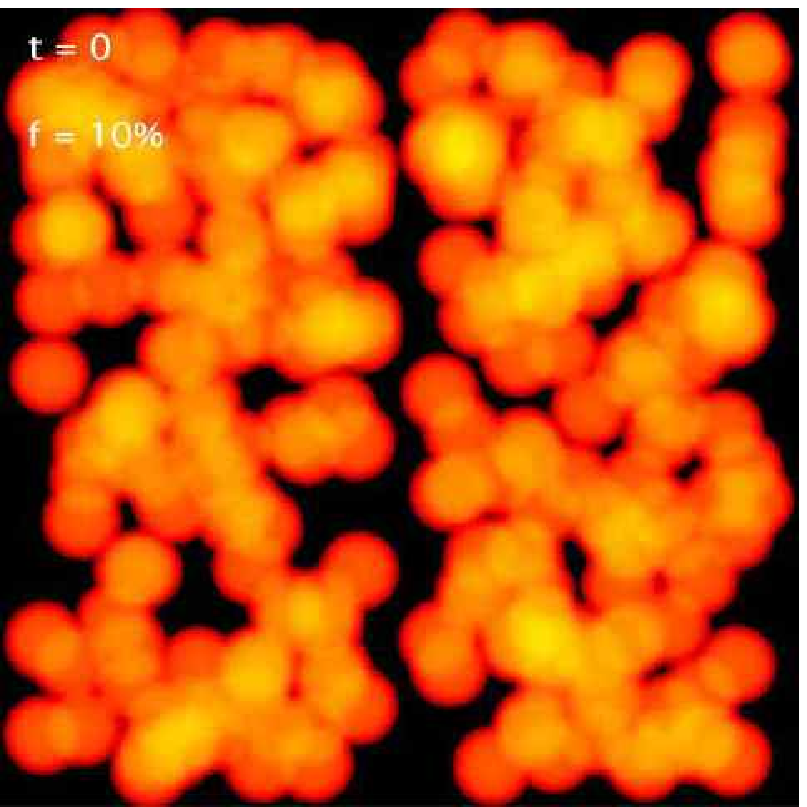}		
			\psfrag{time}{\textcolor{white}{t = $1 / 2$ \tcrpic}}
			\psfrag{fill}{\textcolor{white}{\Vffpic ~= 10\%}}
		\includegraphics[width=2.1in,height=2.1in]{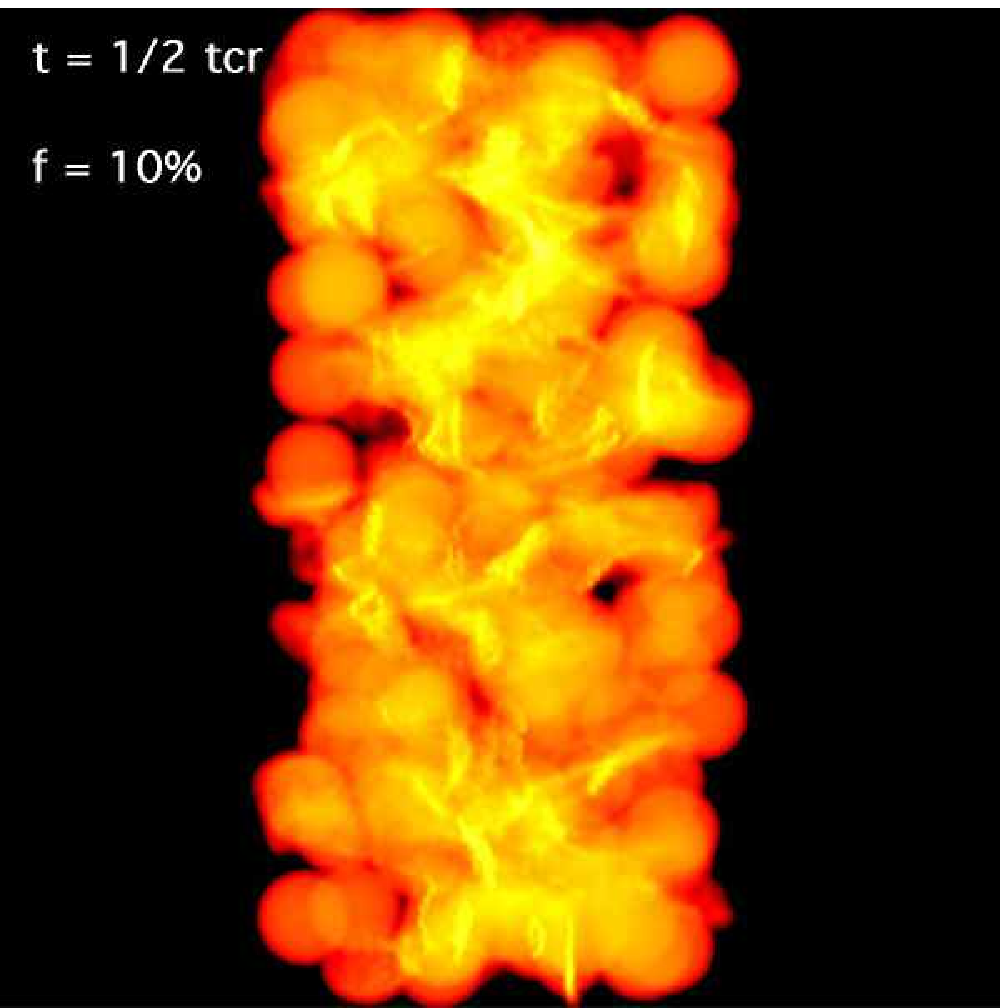}
			\psfrag{time}{\textcolor{white}{t =  \tcrpic}}
			\psfrag{fill}{\textcolor{white}{\Vffpic ~= 10\%}}	
		\includegraphics[width=2.1in,height=2.1in]{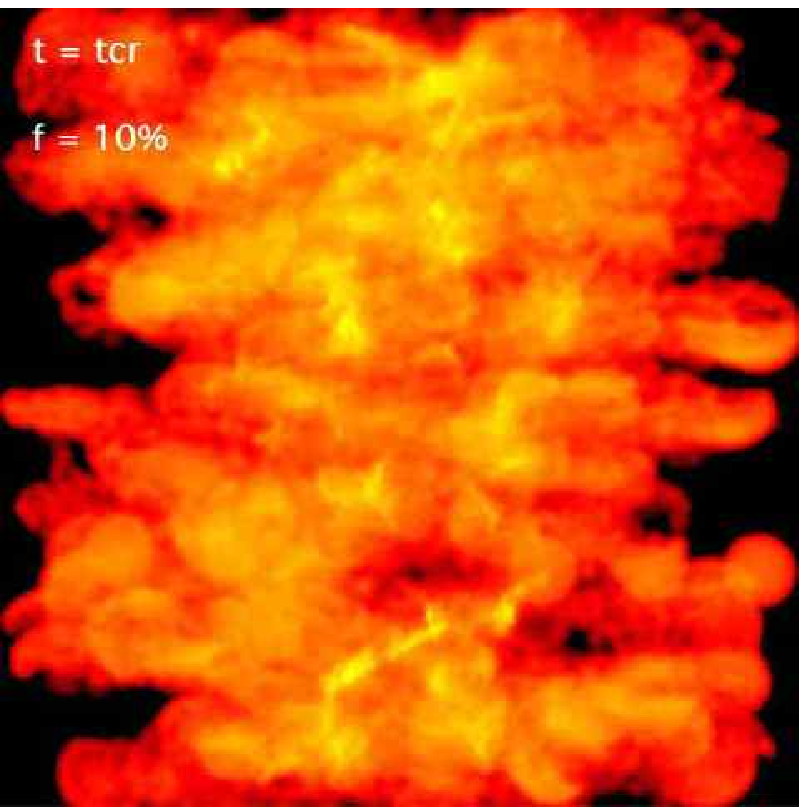}	}		
\centerline{ 	\psfrag{time}{\textcolor{white}{t  = 0.}}
			\psfrag{fill}{\textcolor{white}{\Vffpic ~= 20\%}}
		\includegraphics[width=2.1in,height=2.1in]{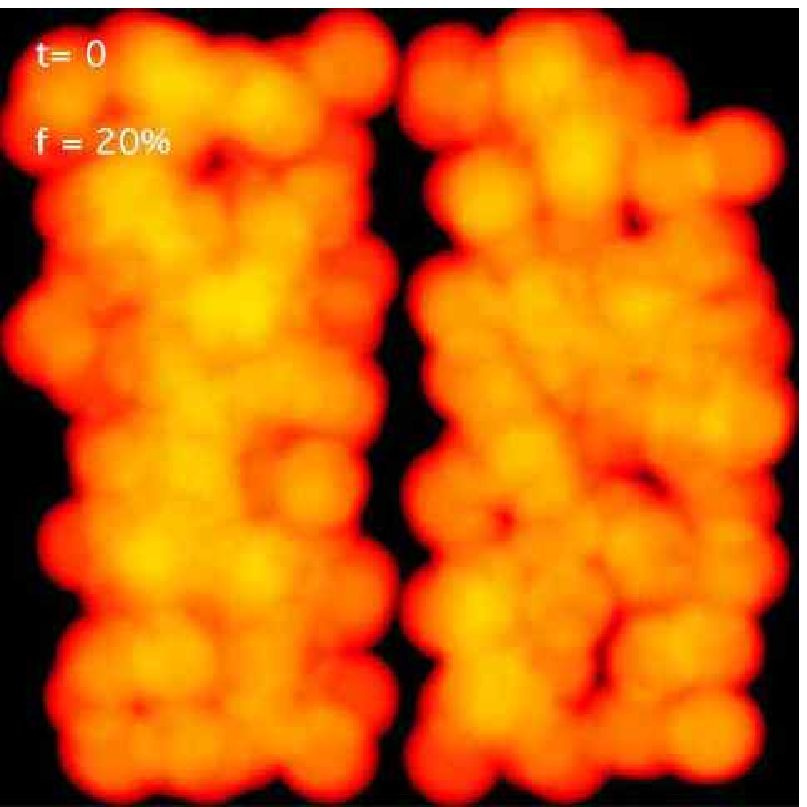}	
			\psfrag{time}{\textcolor{white}{t = $1 / 2$ \tcrpic}}
			\psfrag{fill}{\textcolor{white}{\Vffpic ~= 20\%}}	
		\includegraphics[width=2.1in,height=2.1in]{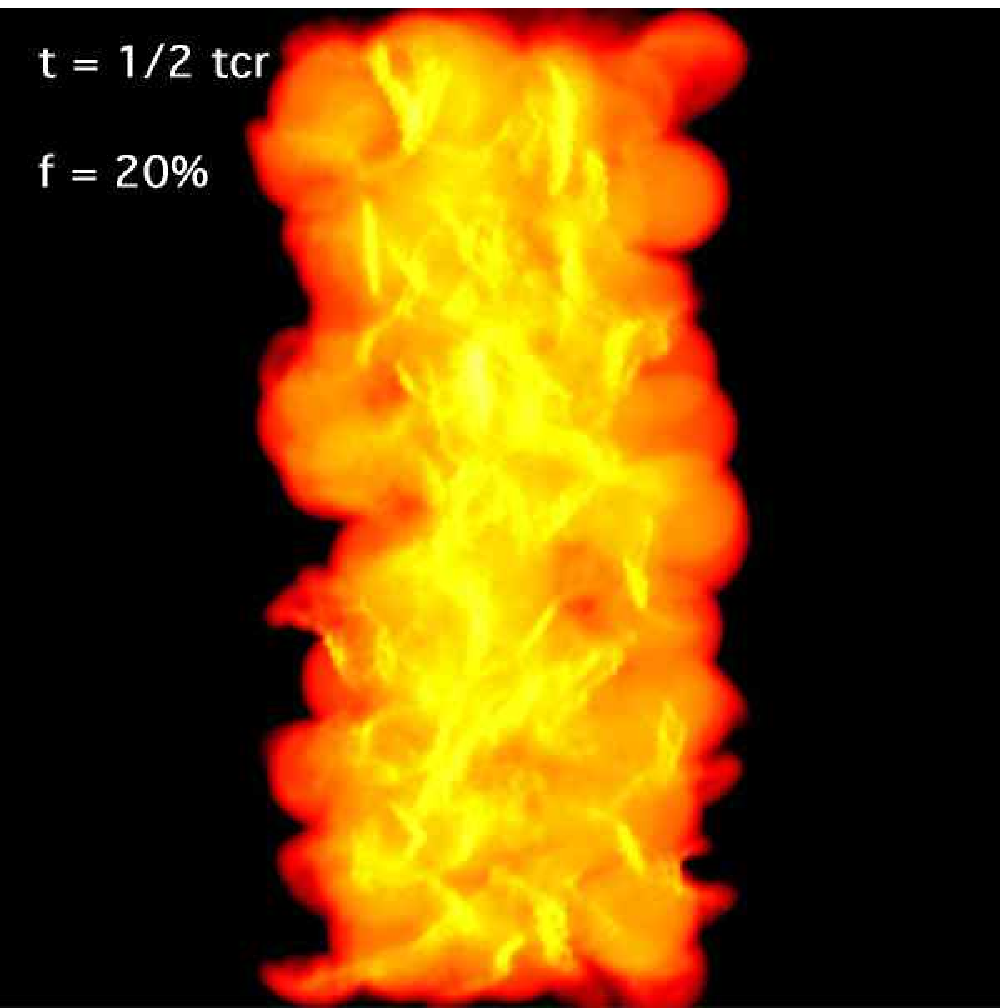}
			\psfrag{time}{\textcolor{white}{t =  \tcrpic}}
			\psfrag{fill}{\textcolor{white}{\Vffpic ~= 20\%}}	
		\includegraphics[width=2.1in,height=2.1in]{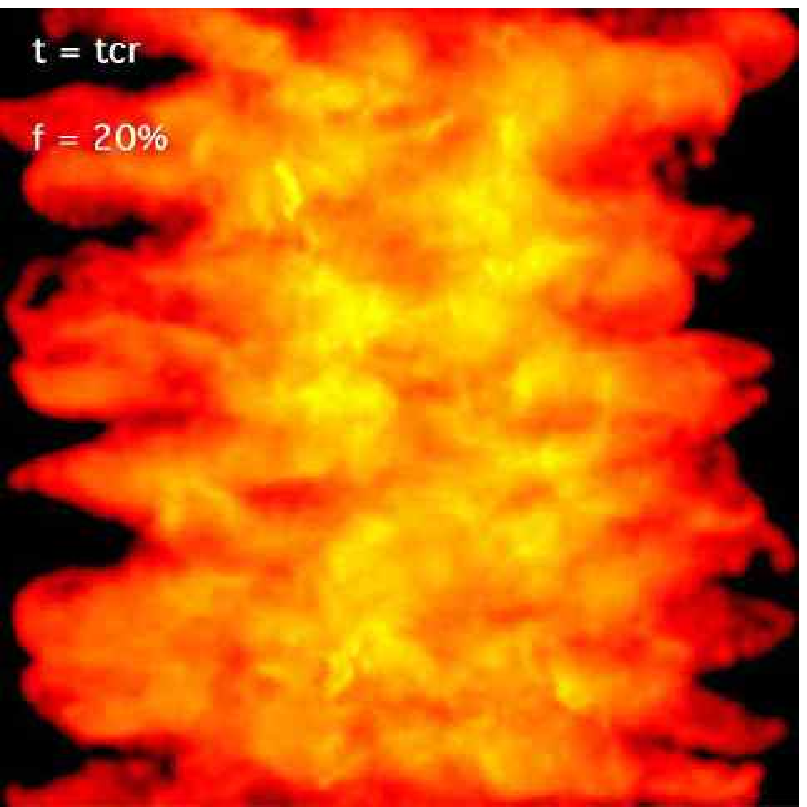}	}	
\centerline{  	\psfrag{time}{\textcolor{white}{t = 0.}}
			\psfrag{fill}{\textcolor{white}{\Vffpic ~= 40\%}}
		\includegraphics[width=2.1in,height=2.1in]{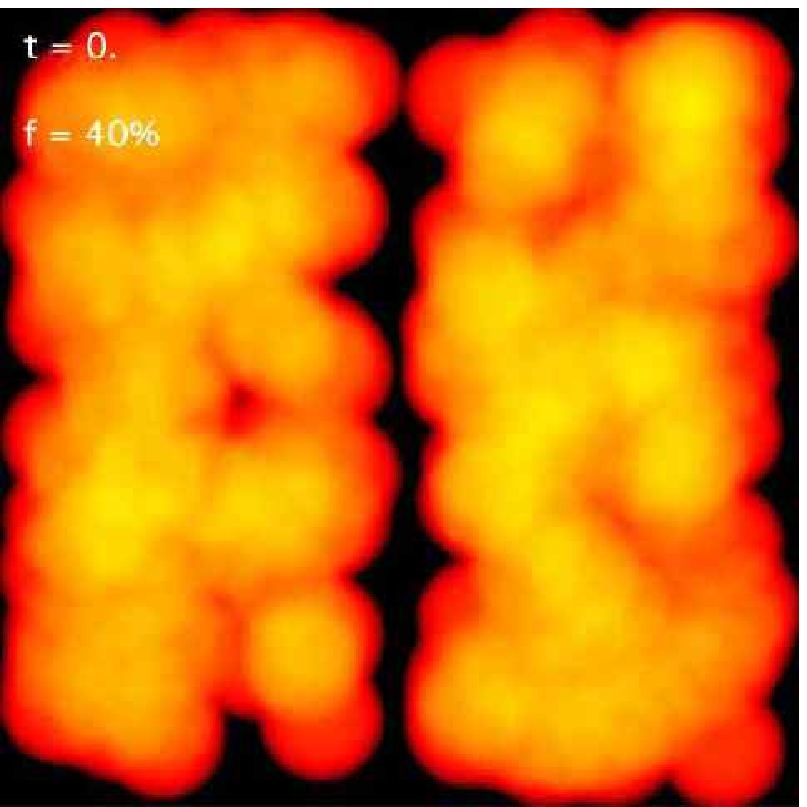}
			\psfrag{time}{\textcolor{white}{t = $1 / 2$ \tcrpic}}
			\psfrag{fill}{\textcolor{white}{\Vffpic ~= 40\%}}	
		\includegraphics[width=2.1in,height=2.1in]{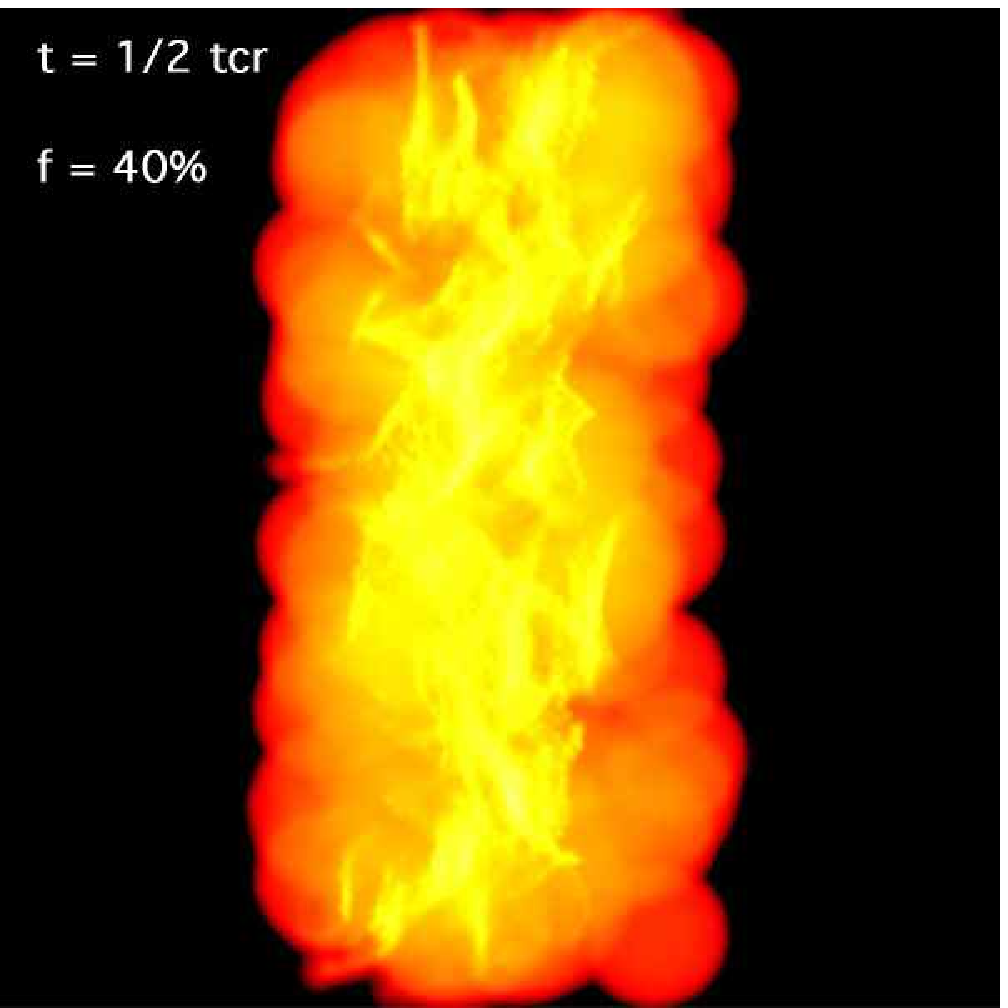}
			\psfrag{time}{\textcolor{white}{t =  \tcrpic}}
			\psfrag{fill}{\textcolor{white}{\Vffpic ~= 40\%}}	
		\includegraphics[width=2.1in,height=2.1in]{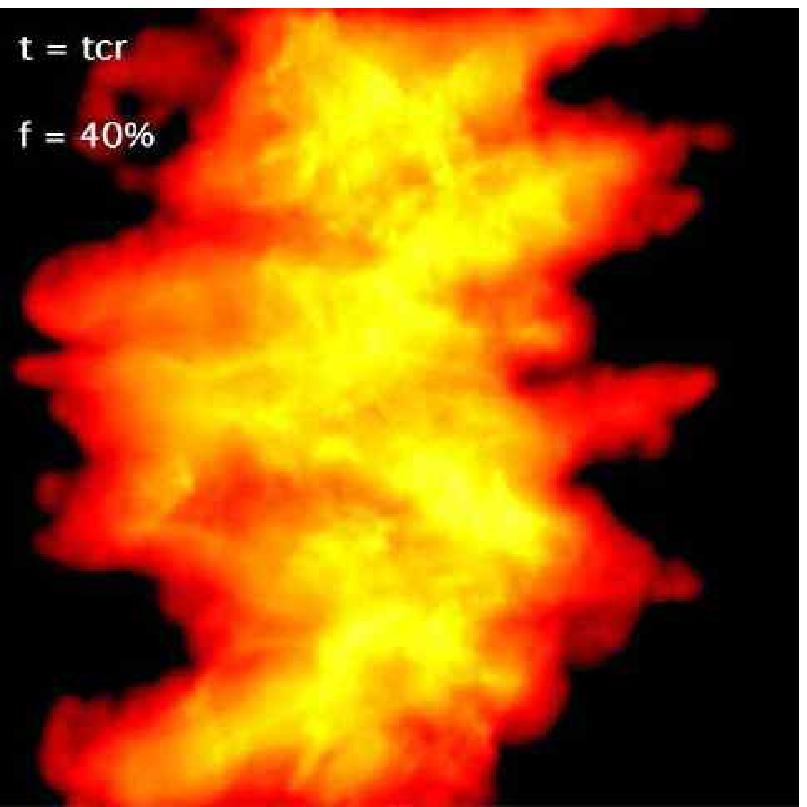}	}
}}
\caption{\label{clumpspicture} We plot here column density images taken from the Mach
10 simulations. The images are taken at a time t = 0, t = 0.5\tcr, and t = ~\tcr, where \tcr~is the
crossing time for the flows, defined as $t_{cr} = L_{flow}/v_{flow}$. For these simulations, 
$L_{flow}$ is just half the length of the region shown. The top, middle and bottom rows 
contain the images from the 10\%, 20\% and 40\% filling factor simulations respectively. 
The colour table for these images is stretched over a range from 0.0001 g cm$^{-2}$ to 10 
g cm$^{-2}$}.
\end{figure*}

\begin{figure*}
\centerline{ 	\includegraphics[width=2.8in,height=2.8in]{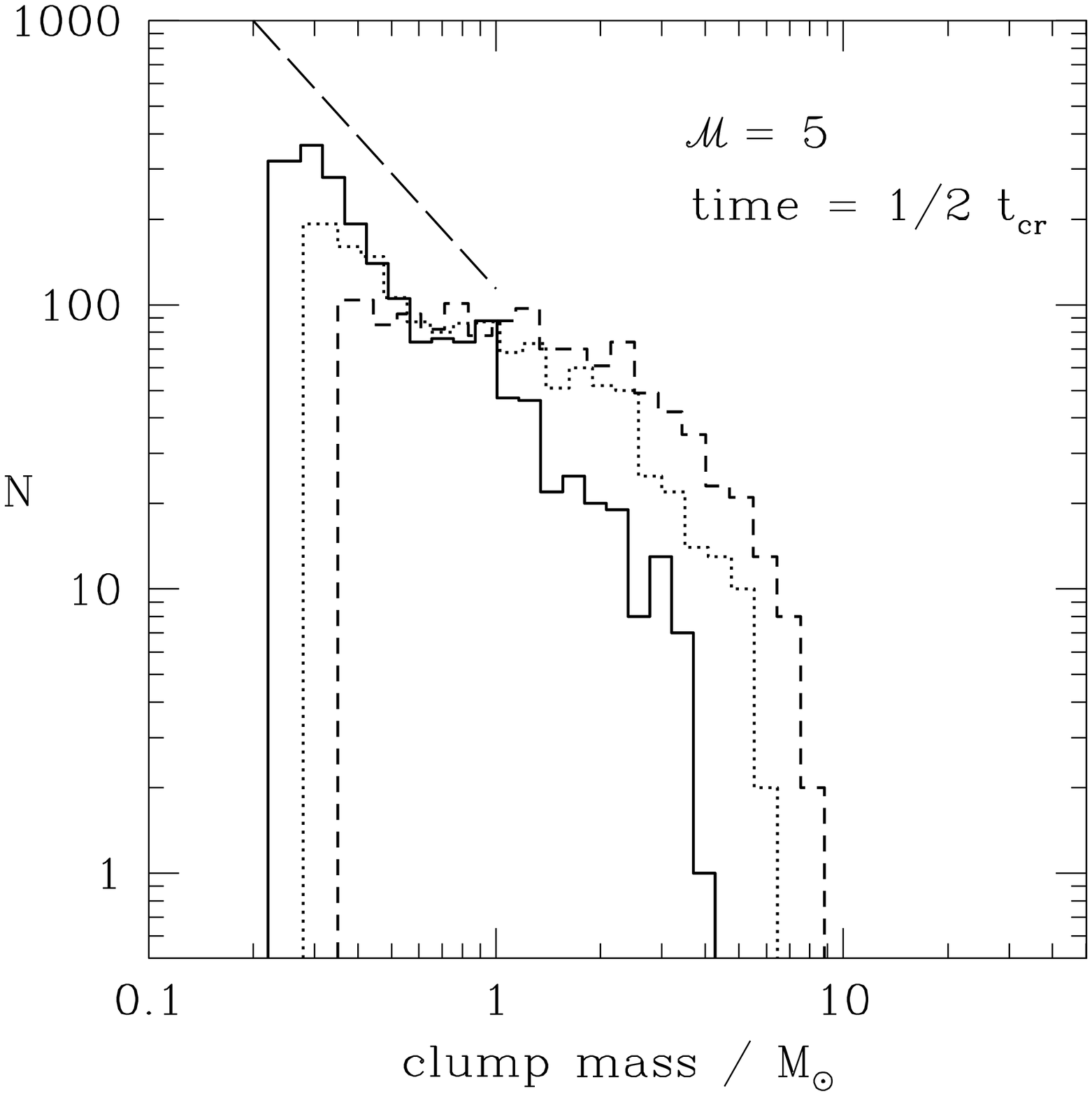}  
			\includegraphics[width=2.8in,height=2.8in]{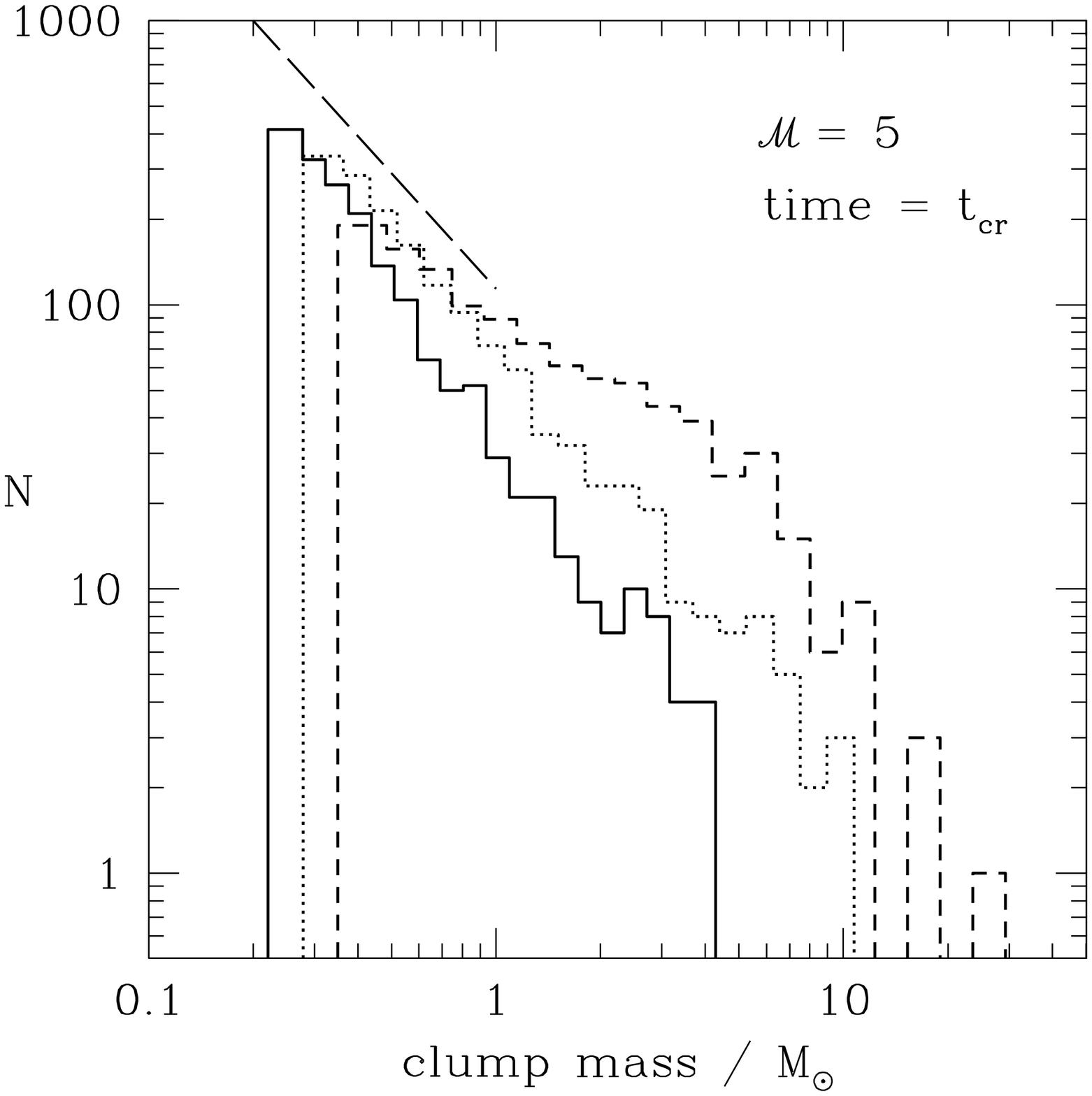}   		}
\centerline{	\includegraphics[width=2.8in,height=2.8in]{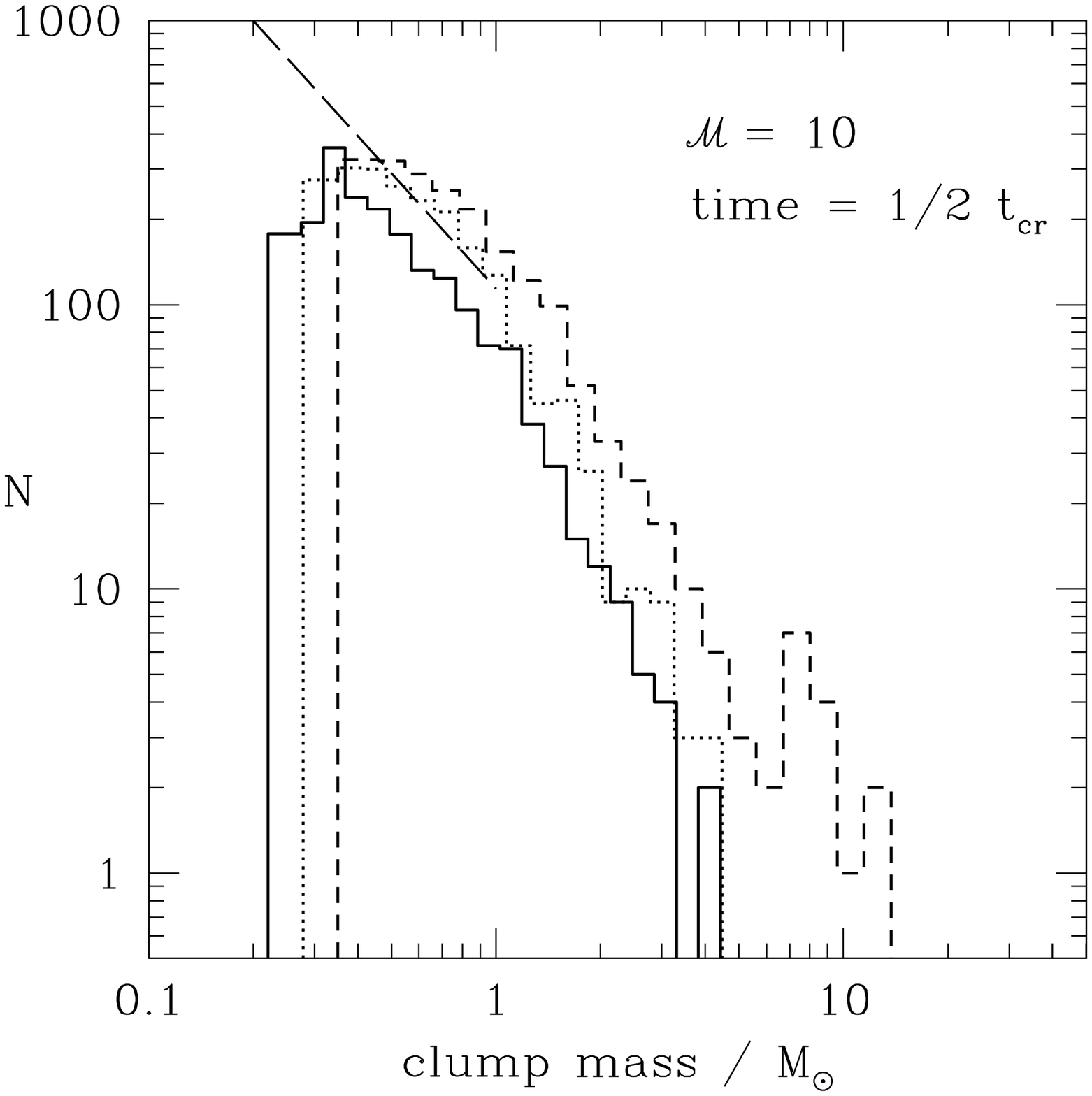} 
			\includegraphics[width=2.8in,height=2.8in]{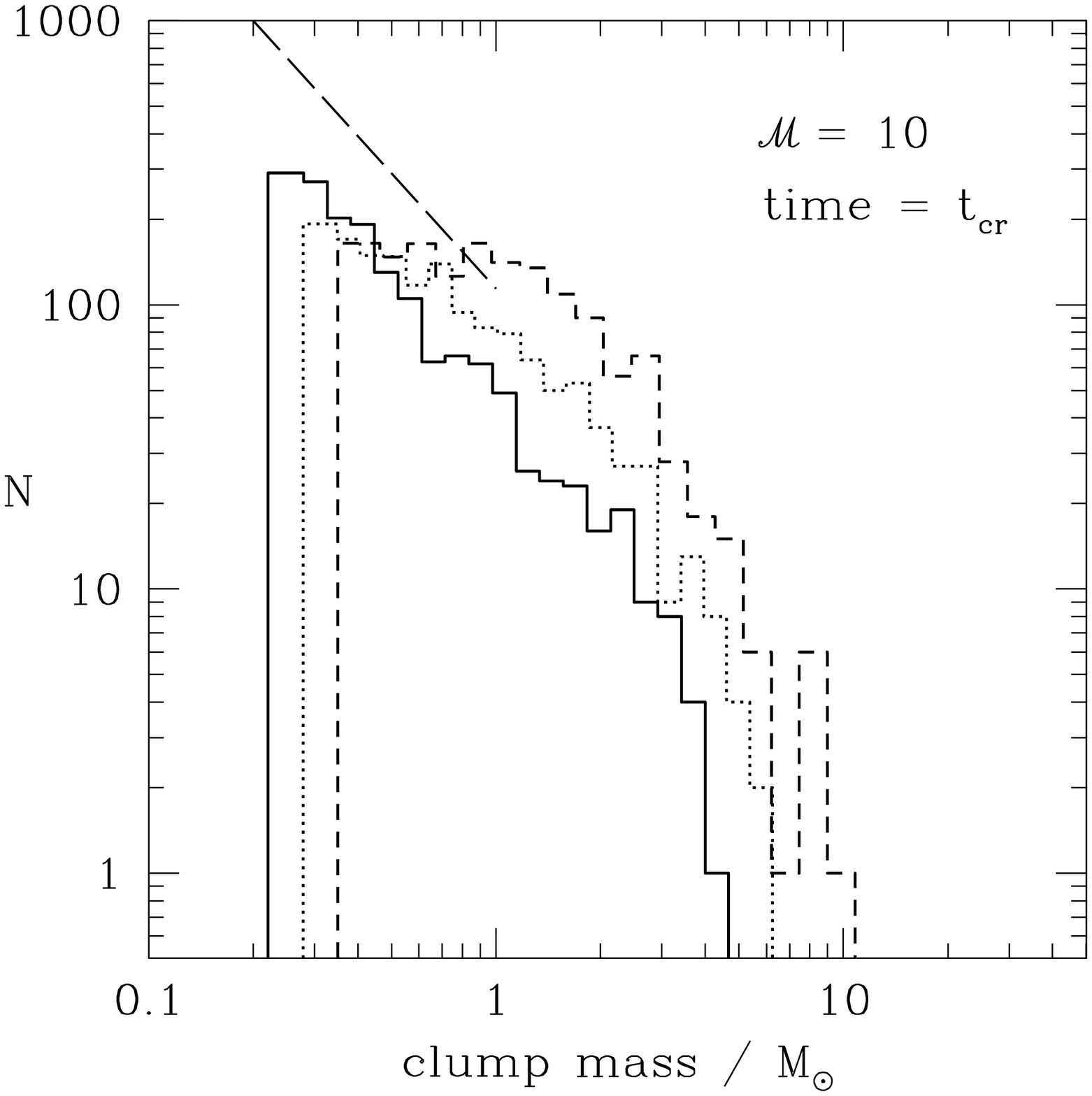} 		} 
\centerline{ 	\includegraphics[width=2.8in,height=2.8in]{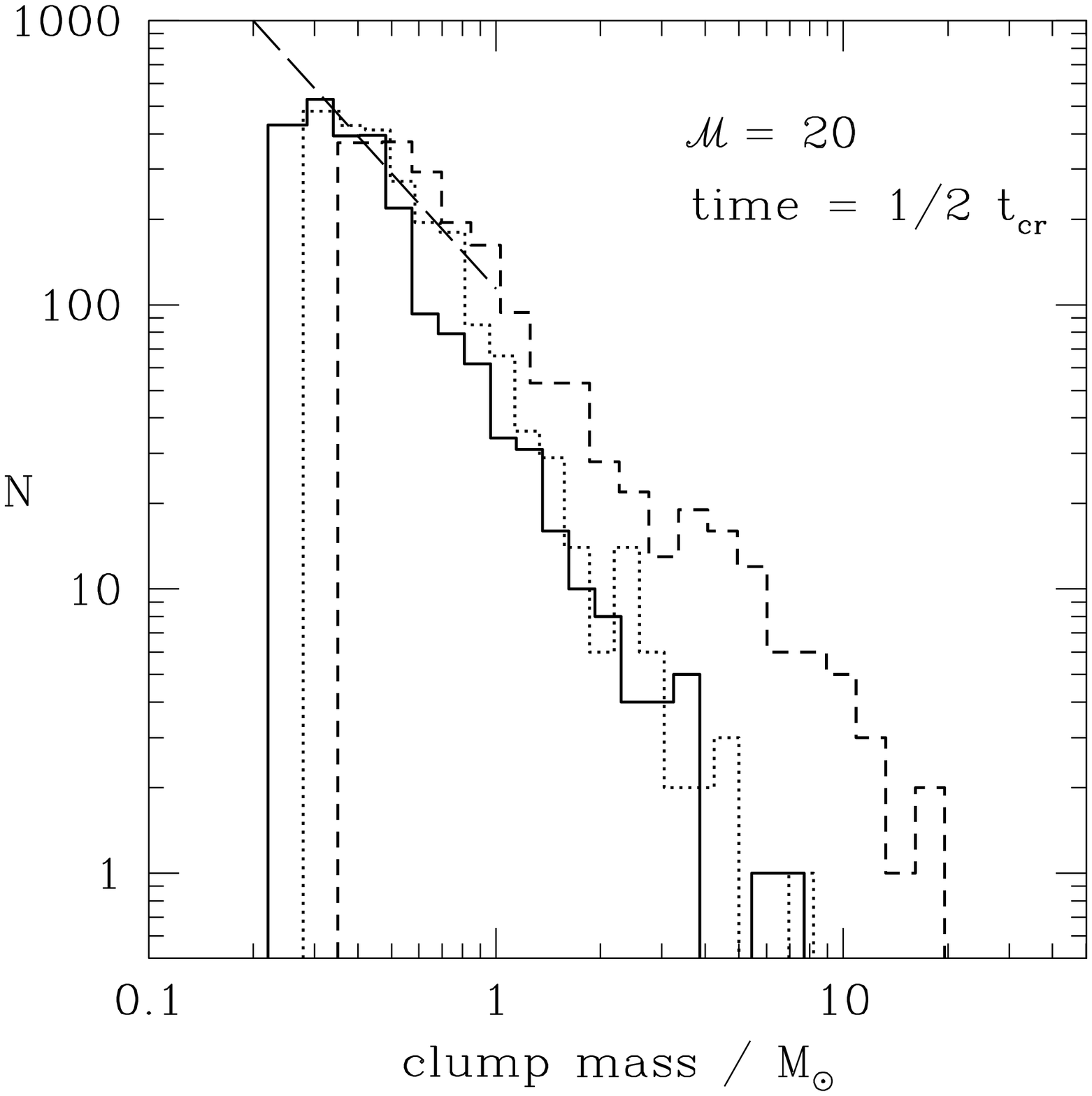} 
			\includegraphics[width=2.8in,height=2.8in]{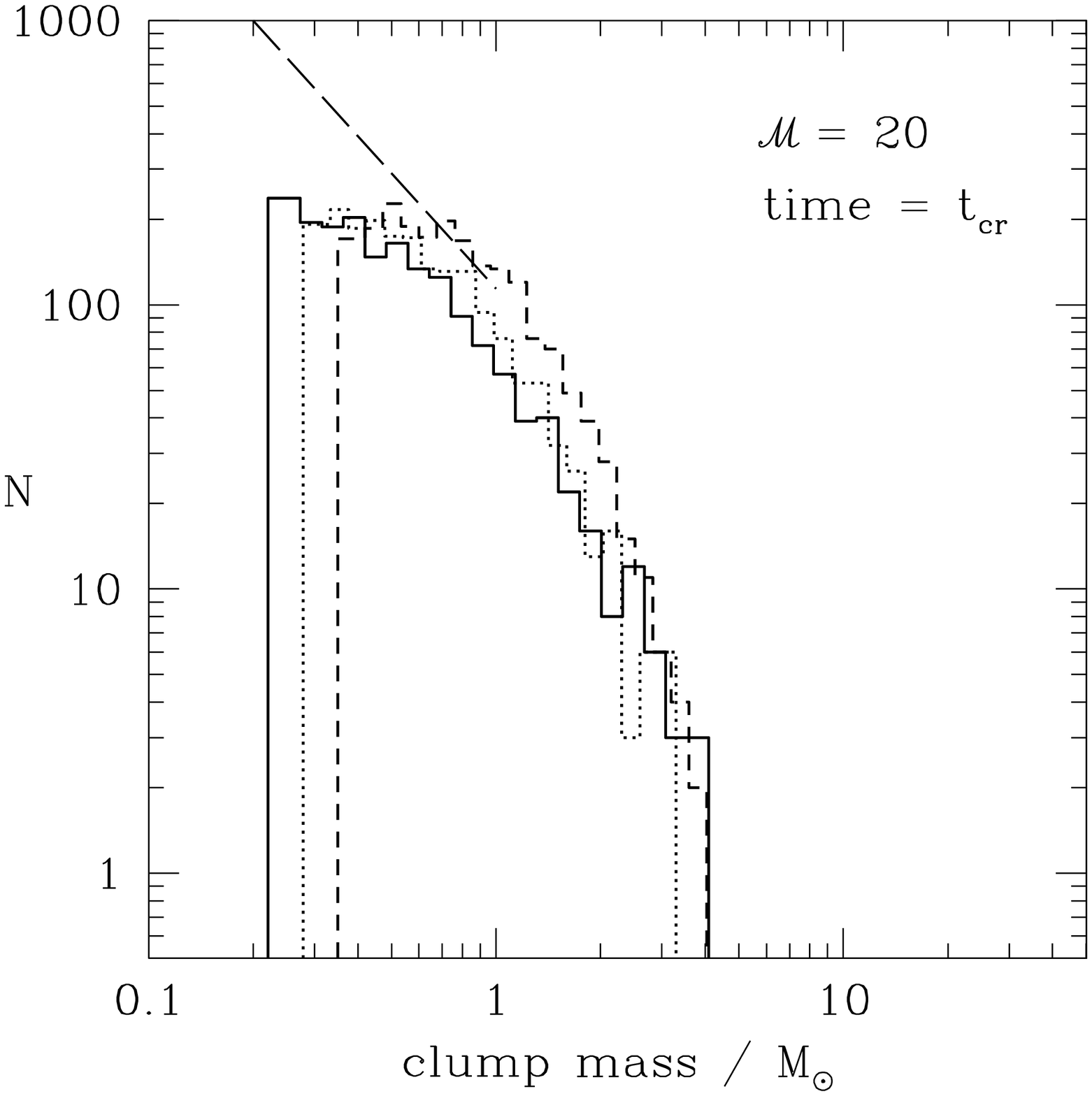} 		}
\caption{\label{clumpspectra} Shown in this figure are the clump mass spectra for
all the calculations {\it{without}} self-gravity. The spectra are shown at two instances
during the flow evolution. On the left, the spectra are taken at a time $t = 0.5$\tcr, 
corresponding to the point where the whole cloud is at its most compact (that is,
the full length of the flows has entered the shock). On the right, we show the spectra at
time $t = $~\tcr, which is the point where the flows have completely emerged from the 
shock. In each of the plots we represent the three filling factors. The solid, dotted
and dashed distributions showing the spectra from the 10, 20 and 40\% filling
factor flows, respectively. The long dashed line shows a Salpeter slope of $\gamma =
2.35$ (so $-1.35$ on our log-log plot here).}
\end{figure*}

%
% Simulation Details...
%

\section{The simulations}
\label{calculations}
\subsection{The fluid code}
\label{code}

The fluid is modelled using the Lagrangian particle method of smoothed
particle hydrodynamics, or SPH \citep{Lucy1977, GingoldMonaghan1977,
Monaghan1992}. The smoothing lengths are variable in both time and space, with
the constraint that there must be a roughly constant number of neighbours for
each particle, which is chosen to be roughly 50 (with a fluctuation from 30 to
70 neighbours). We use the standard artificial viscosity suggested by
\citet{MonaghanGingold1983} with an alpha and beta of 1 and 2 respectively.
Gravitational forces are calculated using quadrupole moments obtained via a
tree structure \citep{Benzetal1990}, which is also used to construct particle
neighbour lists.

The code includes the modification by \citet*{Bateetal1995} such that dense,
bound, regions of gas are replaced by point masses, or `sink particles'. The
sinks interact only via gravity but are able to accrete gas particles within
their accretion radius. These sinks allow the code to model the dynamical
evolution of accreting protostars, without integration times becoming
prohibitively small. Only three of the simulations presented here include gravitational
forces and sink particles. The detail of the sink particle formation in these three
simulations in given below.

\subsection{Initial conditions}
\label{setup}

In this paper, we present a series of simulations following the evolution of colliding
flows, in which the flows consist of a population of clumps. The clumps are created
by populating the computational volume with a settled template clump of 2000
particles. The clump radii are then set to provide a desired filling factor. Masses of
the clumps are then determined by assuming each template clump has a Jeans 
number of $J$ = 0.5 at a temperature of 10K. The Mach number of flows is also changed
in the simulations, from Mach 5 to Mach 20. A summary of the initial conditions for each 
of the simulations can be found in Table \ref{initialdetails}.

We now go discuss this process of constructing the flows in more detail.
The clumpy flows are formed by populating the computational volume with copies of 
a template clump. To construct the original template clump, we place 2000 SPH 
particles randomly in a spherical volume. The particles are given no initial velocities. 
We then evolve this clump in isolation for ten dynamical times\footnote{The dynamical
time is just the sound crossing time for the gas sphere.} with a spherical, constant volume, 
boundary condition. This allows the SPH particles to settle into a uniform density 
sphere and gives time for particle velocities, which result from the random initial positions, 
to be damped by the artificial viscosity. 
When the template clump is used to create the clumpy flow, each copy is 
given a random orientation before being placed in the computational volume.

Two properties of the flow are altered in this paper: the Mach number and the filling 
factor. The number of initial clumps is held constant at 250, and the volume of the 
flow is kept the same, with the computational volume being a cube of side 8 pc. To 
alter the filling factor, the volume of the clumps is then changed such that the clumps 
occupy the desired fraction of the computational volume. We also impose the constraint 
that the clumps are unbound by their internal energy, with a Jeans number 
$J = |E_{grav}|/E_{therm} = 0.5$. They clumps are also given a temperature of 10K, which 
remains constant throughout the simulation, since we assume an isothermal equation
of state. The combination of filling factor and Jeans number then sets the required initial 
mass, and thus density of the clumps. 
Without some kind of external pressure, the 
clumps, being unbound, would just expand and eventually merge. To prevent this
occurring, we set an external pressure term which is equal to the thermal pressure
of the clump gas. 

The flow velocities are set up along the $x$ axis, such that particles with $x < 0$ have a 
velocity $v_{x} = +v_{flow} = +{\mathcal{M}}c_{s}$ and particles with $x > 0$ have a velocity
$v_{x} = -v_{flow} = -{\mathcal{M}}c_{s}$. The $y$ and $z$ components of the velocity are 
set to 0. As mentioned above, the cubes have a side of 8 pc in our simulations, which goes 
from $-4$ to 4 pc in each of the $x$, $y$ and $z$ directions. In this paper we will refer to the 
{\it{crossing time}}, which we will denote by \tcr. The crossing time is given by 
$t_{cr} = L_{flow}/v_{flow}$. Since our flows only take up half the length of the computational 
volume, the crossing time is $t_{cr} = 4$ pc $/ ({\mathcal{M}}c_{s})$. We define the crossing 
time in this manner so that the shock is finished at t = \tcr.

In this paper, three different flow velocities are presented. These are Mach 5, 10 and 20,
which with a sound speed of 0.2 km s$^{-1}$, gives flow speeds of 1, 2 and 4 km s$^{-1}$,
respectively. The corresponding crossing times for the Mach numbers 5, 10
and 20 are 4, 2 and 1 Myrs, respectively.
For each of these Mach numbers, we also simulate 3 different filling factors, with the 
clumps occupying 10, 20 and 40\% of the flows. The filling factors will be denoted in this
paper by the symbol \Vff.
The template clump which is used to populate the flows has a mass
of 4.4 \solmasp, 5.6 \solmas and 7 \solmas for the 10\%, 20\% and 40\% filling factor
simulations, respectively.
There are thus 9 different initial conditions 
for the simulations. In these simulations we do not include gravitational forces, since our 
first aim is examine the ability of gas to form clump-mass spectra purely through collisions. 
However for the simulations discussed in section \ref{gravsection}, the self-gravity 
is switched on, allowing us to examine 
the effect of the gravity on the clump mass spectra, and the formation of stars.

To follow the star formation in simulations which include self-gravity, 
we create `sink particles' when the 
density reaches a certain limit. In SPH calculations involving self-gravity, there is a limiting 
mass resolution below which the SPH method will not reliably form self-gravitating objects
\citep{BateBurkert1997}. \citet{BBB2003} showed that this occurs when the objects are 
represented by less than roughly 75 SPH particles. The mass resolution of any SPH 
simulation can thus be
given by, $m_{res} \sim 100(M_{sim}/N_{part})$. In our calculations, we create the sink 
particles at the point where $m_{res}$ becomes self-gravitating, and thus has a Jeans mass. 
The Jeans mass is given by, 
\begin{equation}
\label{jeansmass} 
m_{J} = \Big(\frac{4 \pi \rho}{3}\Big)^{-1/2}
\Big(\frac{5}{2}\frac{kT}{G\mu}\Big)^{3/2} 
\end{equation}
and one can rearrange for the density to obtain the density threshold above which sink 
particles can be created. The mass resolution for our simulations which include 
self-gravity (with \Vff~= 40\%) is \sims 0.3 \solmas, 
which gives a sink particle creation density of $2.47 \times 10^{-18}$ g cm$^{-3}$. The 
radius associated with this mass resolution and density is $3.87 \times 10^{16}$ cm 
(or 0.012 pc) and represents the accretion radius for the sink particles, inside which 
they can accrete gas particles. To ensure that our codes run quickly, we smooth the 
gravitational forces between sink particles to the accretion radius. This prevents binaries 
with separations smaller than 0.012 pc from forming in our simulations, since they are 
not of interest to this paper.

\subsection{Finding the clumps} 
\label{structure} 

The main aim of this paper is to examine the formation of clump populations in colliding
flows. This study uses a clump finding algorithm based on the method developed by
\citet*{KlessenBurkert2000}. This is in turn based on the method developed by 
\citet*{Williamsetal1994}. We point the interested reader to the Appendix 1 in
\citet*{KlessenBurkert2000} for a description of the method. This algorithm
makes use of the SPH formalism and has been shown to very easily identify
density structure in the gas. These structures are analysed in this paper in
order to assess their role in the star formation process.

However unlike both \citet{Williamsetal1994} and \citet*{KlessenBurkert2000},
we do not use a density threshold to define what is termed as a `clump', i.e.
there is no condition on how much higher the peak density has to be in relation
to the minimum density of the structure. While this means that, technically,
very low contrast density features are free to be defined as clumps, in
practise this never occurs in our simulations. The clumps, formed from
supersonic flows, have a density peaks that are an order of magnitude higher
than the mean gas density.

%
% Clump mass spectra ---> No Gravity
%

\section{The clump mass spectra}
\label{nogravsection}

In Figure \ref{clumpspicture}, we show column density images for the simulations
in which the flow has a Mach number of 10. Each of the three filling factors that is
investigating in this study are presented, and images are shown at three times
during the simulations: first the initial flow setup, then at $t = 0.5$\tcr (half crossing 
time, when the whole ensemble of gas is at its most compact), and at $t = $\tcr, when
the flows are roughly back to their original size. One can easily see that at the half
crossing time, the cloud is compact, and that it is heavily structured with dense 
filaments that have formed as the colliding clumps produce shocks. After a crossing
time, the cloud has less shock structure, and the density features are more rounded,
since the shock filaments have had time to re-expand. The region now has a mottled
and wispy appearance, similar to what one observes in the gas structure around
embedded clusters. 

The purpose of this study is to see whether the process of creating
a dense region from a clumpy flow, can create a population of clumps in the region
that resemble the mass spectrum for stars.
The clump mass spectra are shown in Figure \ref{clumpspectra}. This figure shows the
mass spectra for each of the simulations, taken at $t = 0.5$\tcr~and $t = $\tcr. Note that the
clumps shown in the figure, as found by the clump-finding algorithm (see Section 
\ref{structure}), are 3-dimensional.

It is clear from the plots in Figure \ref{clumpspectra} that the clump mass 
spectra evolve as the flows enter the shock. The distributions at $t = 0.5$\tcr~and 
$t = $\tcr~look distinctly different. The amount of evolution in the spectra depends on 
the volume filling factor \Vff, with the lower \Vff~simulations changing much less than 
those with higher values of \Vff. This result is not that surprising. In the calculations 
with the high filling factors, there is a much greater chance of two clumps colliding
and potentially shredding.

The Mach number of the flows can also have an effect on the mass distribution of
clumps. In the Mach 5 simulations, there are more high mass objects at $t = $\tcr than
there are at $t = $\tcr.  The distributions are also better fitted by a power law. For the
simulations with Mach 10 \& 20 flows, the opposite is true.  The more massive clumps
appear at $t = $0.5\tcr. The distributions is also better fitted by a power law at $t = 0.5$\tcr,
and by comparison, the clumps are less massive at $t = $\tcr, and not easily fitted by
a power law.

This dependence on the Mach number for the clump mass spectra suggests that
two different processes are at play here, coagulation and shredding, and that the
efficiency of these processes is linked to the strength of the shocks. One thing that
is certain is that the shredding is always much more efficient here than the 
coagulation. The initial clump that is used to populate the flows has a mass of 
roughly 4 to 7\solmas, depending on the filling factor of the flow in question, and there
are only 250 clumps present initially, so the shredding process clearly dominates.

At lower Mach numbers however (represented here
by our Mach 5 simulations), coagulation does seem to be present. Low
mass clumps gradually grow through merging events with more massive clumps. 
The low mass clumps must therefore survive for long enough (that is, no re-expansion) to 
be able to keep merging with other clumps. For the clumps to get progressively more 
massive as the flows interact, implies that either the time scale for their 
re-expansion is less than \tcr, or that the clumps are in equilibrium with the pressure
that we included in the simulations. 

At higher Mach numbers, there appears to be no such steady coagulation process. 
In the Mach 10 flows, the clump mass spectra don't appear to undergo much
evolution between times 0.5\tcr and \tcr. This suggests that the balance between
shredding and coagulation, whatever it may be, is fairly constant throughout the
simulations. It is interesting to note here that only the 40\% filling factor flows manage
to produce clumps that are more massive than those at the start of the simulation. When
one moves to the Mach 20 simulations, the situation is slightly different. At $t = 0.5$\tcr, the
mass spectra look quite similar to those in the Mach 10 simulations, however the Mach
20 flows manage to form some more massive objects. At $t = $\tcr however, much of then
higher mass structures have been lost. Either, these structures were not in equilibrium with
their surroundings, and re-expanded to become part of the surrounding objects, or
they have been broken up by destructive encounters with other clumps. In either
case, shredding and coagulation are not balanced in the Mach 20
simulations, and the higher mass features in the clump mass spectrum at 
$t = 0.5$\tcr are only transient.

As mentioned above, the purpose here is to determine whether the formation
mechanism for forming regions of star formation is potentially the origin of the
clump mass distributions within. We can see from from Figure \ref{clumpspectra}, 
that many of the clump mass spectra can be fitted by a Salpeter type power law.
It is therefore possible for colliding, clumpy, flows to produce
a clump/core population with a mass spectrum consistent with that for the field stars.
Clearly, from the spectra shown in Figure \ref{clumpspectra}, not all the simulations
display this property at all times, but most do exhibit a Salpeter type slope at some
point.

It is important to note here that none of the resulting clumps will be the sites of star 
formation in these simulations, since the model contains no self-gravity, and gravity has 
played no part in producing the mass spectrum. All the structures represented in Figures
\ref{clumpspicture} and  \ref{clumpspectra} are purely due to the coagulation and shredding 
processes that occur as the flows interact.

These result suggest that the observed clump mass spectra in a number of star forming regions of 
may just be a result of how the region was formed, and not directly part of the collapse and fragmentation process. 

Although many of the more massive objects are lost in the higher mach number flows,
and the mass spectrum looks less like that for field stars by the time $t = $\tcr,
this is not a problem for the model.
Star formation only occurs in the regions of highest density ($\ga10^{5}$cm$^{-3}$) 
within molecular clouds, and, consequently, it is in these regions where the
clump/core observations are made.
In terms of our simulations, the density is highest, and the
cores much better defined, around the time $t \sim$ 0.5\tcr, and it is at this point in the
evolution where the clump mass spectra are most similar to the stellar IMF.

%
% Clump mass spectra ---> With Gravity
%

\begin{figure*}
\normalsize
{\bf{
\centerline{ 	\psfrag{mach}{\textcolor{white}{\Machpic ~= 5}}
			\psfrag{time}{\textcolor{white}{t =  \tcrpic}}
			\psfrag{fillfac}{\textcolor{white}{\Vffpic ~= 40\%}}
		\includegraphics[width=2.3in,height=2.3in]{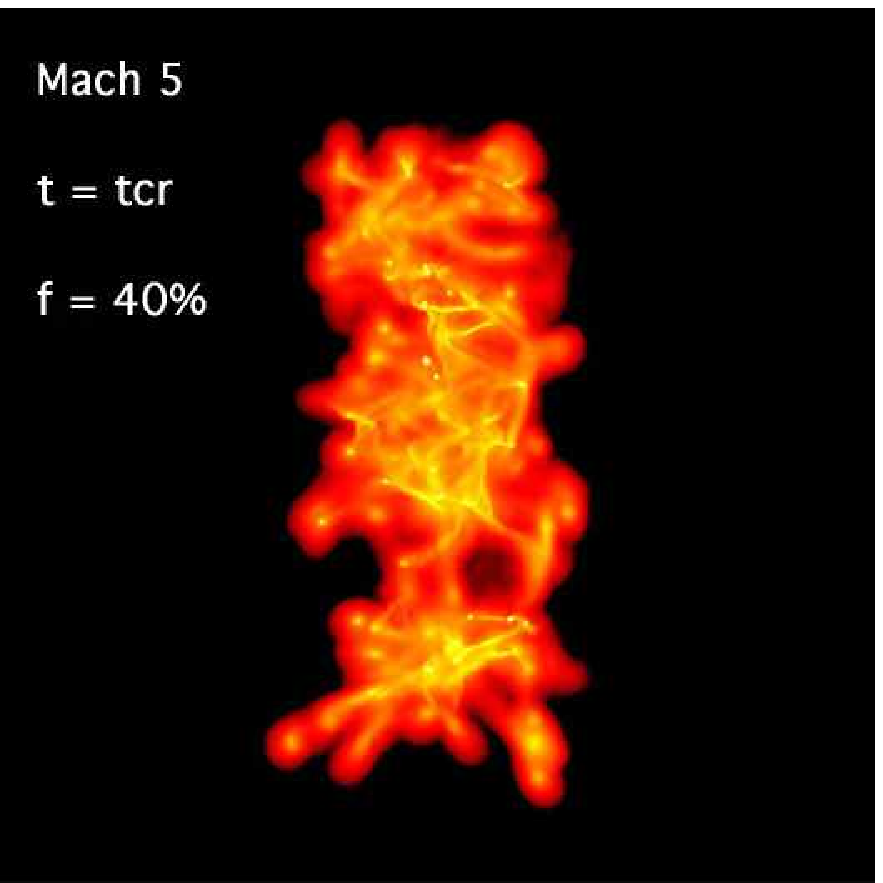}	
			\psfrag{mach}{\textcolor{white}{\Machpic ~= 10}}	
			\psfrag{time}{\textcolor{white}{t =  \tcrpic}}
			\psfrag{fillfac}{\textcolor{white}{\Vffpic ~= 40\%}}
		\includegraphics[width=2.3in,height=2.3in]{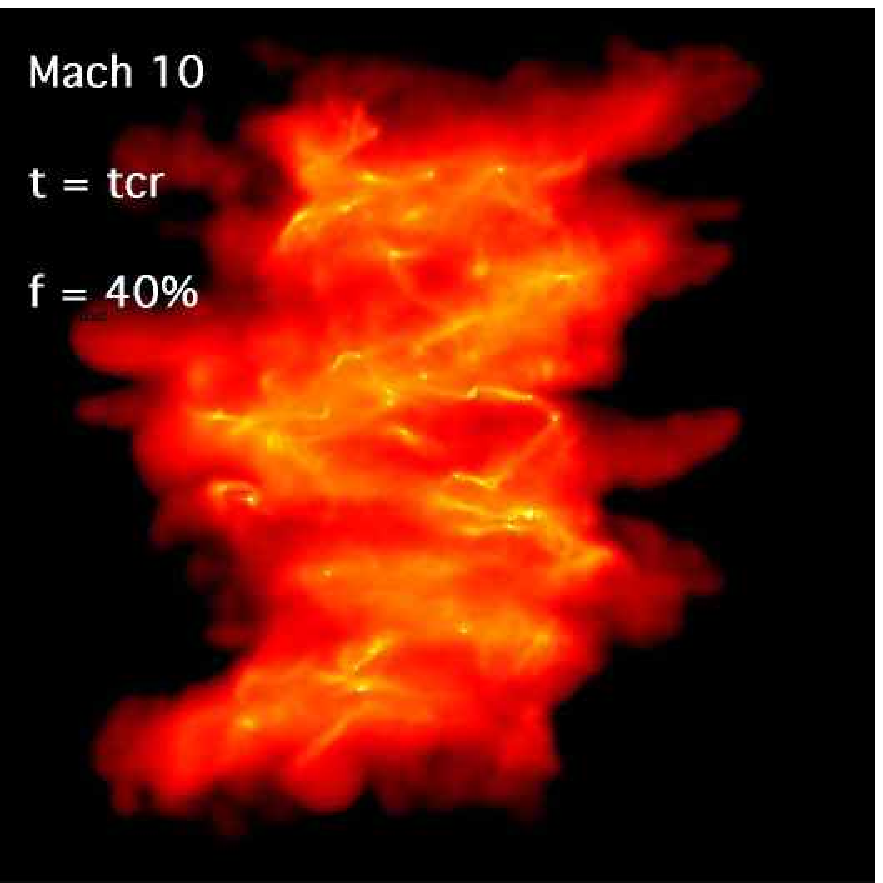}
			\psfrag{mach}{\textcolor{white}{\Machpic ~= 20}}
			\psfrag{time}{\textcolor{white}{t =  \tcrpic}}
			\psfrag{fillfac}{\textcolor{white}{\Vffpic ~= 40\%}}	
		\includegraphics[width=2.3in,height=2.3in]{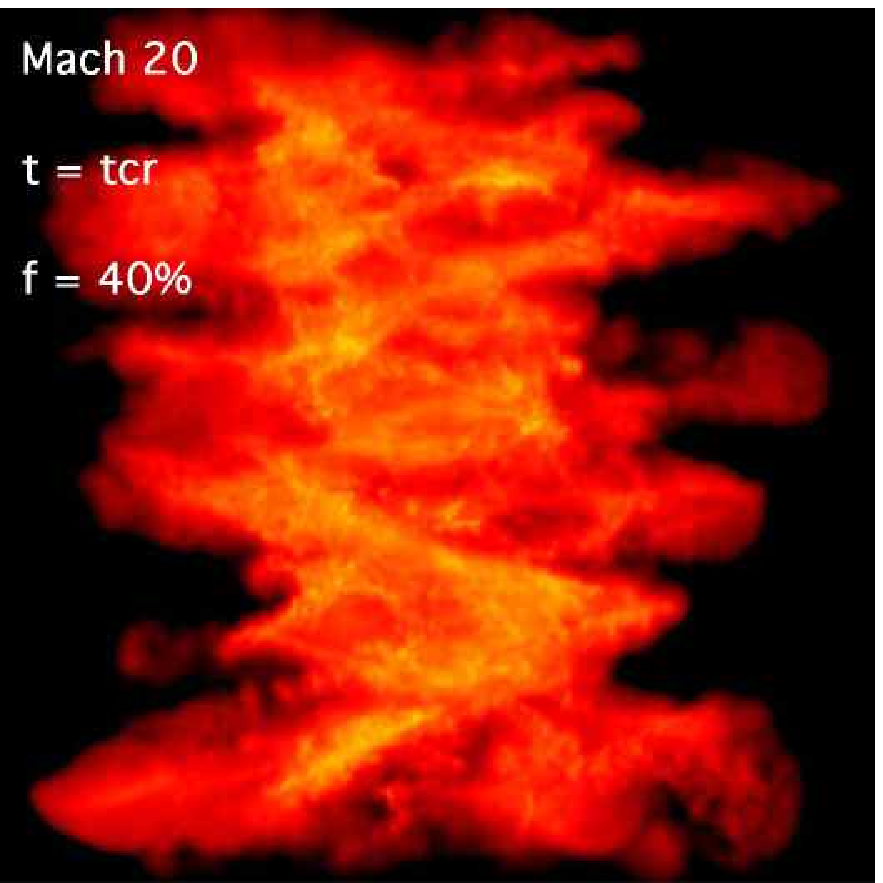}	}
}}
\caption{\label{gravpics} Shown here the column density images of the 40\% filling factor
simulations, this time re-run with the inclusion of self-gravity. The images are taken at the
crossing time for the flow, the point at which the simulations were terminated. The maximum
and minimum column densities plotted are 10 and 0.0001 g cm$^{-2}$, respectively. }
\end{figure*}

\begin{figure*}
\centerline{ 	\includegraphics[width=2.8in,height=2.8in]{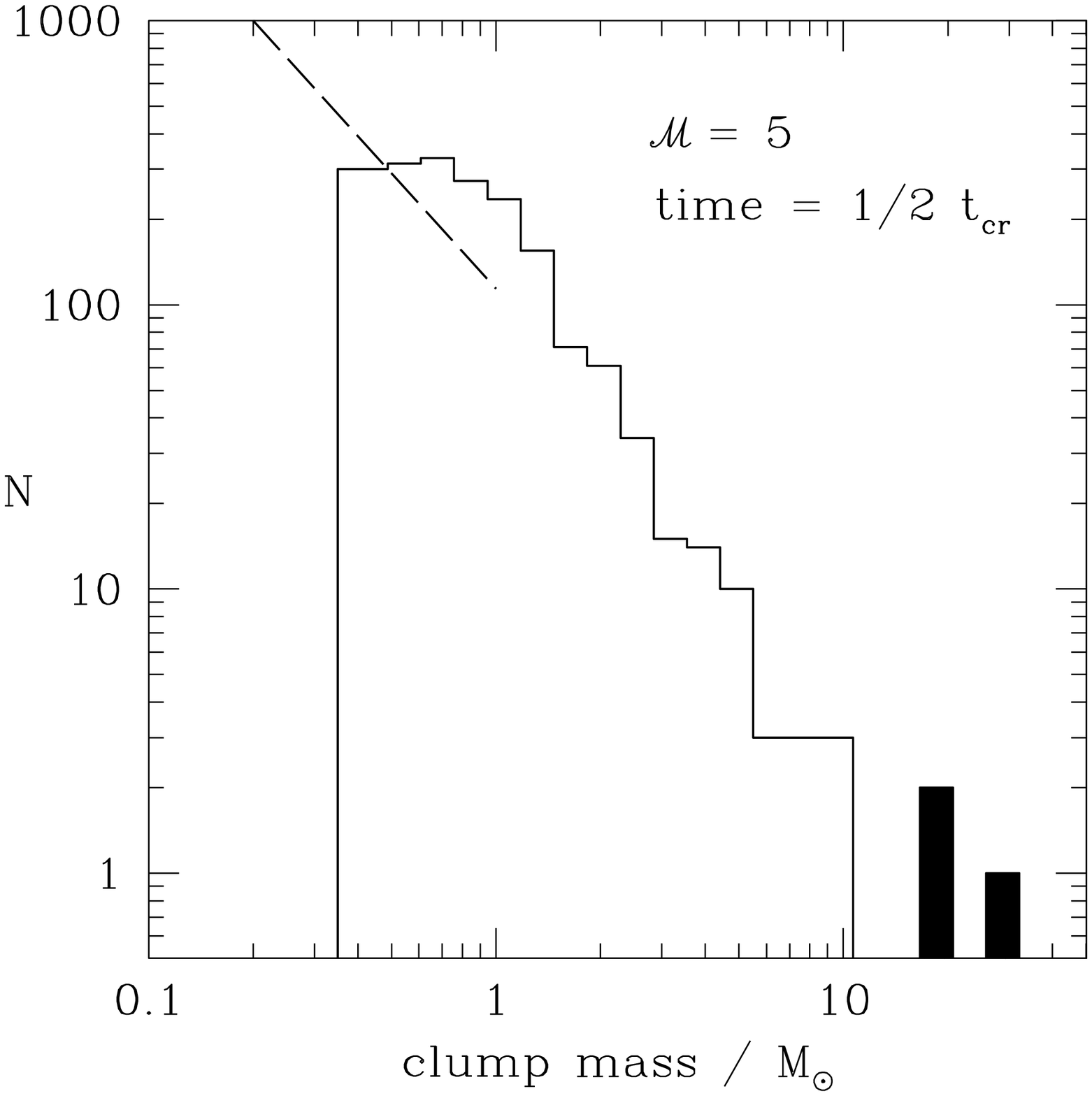}  
			\includegraphics[width=2.8in,height=2.8in]{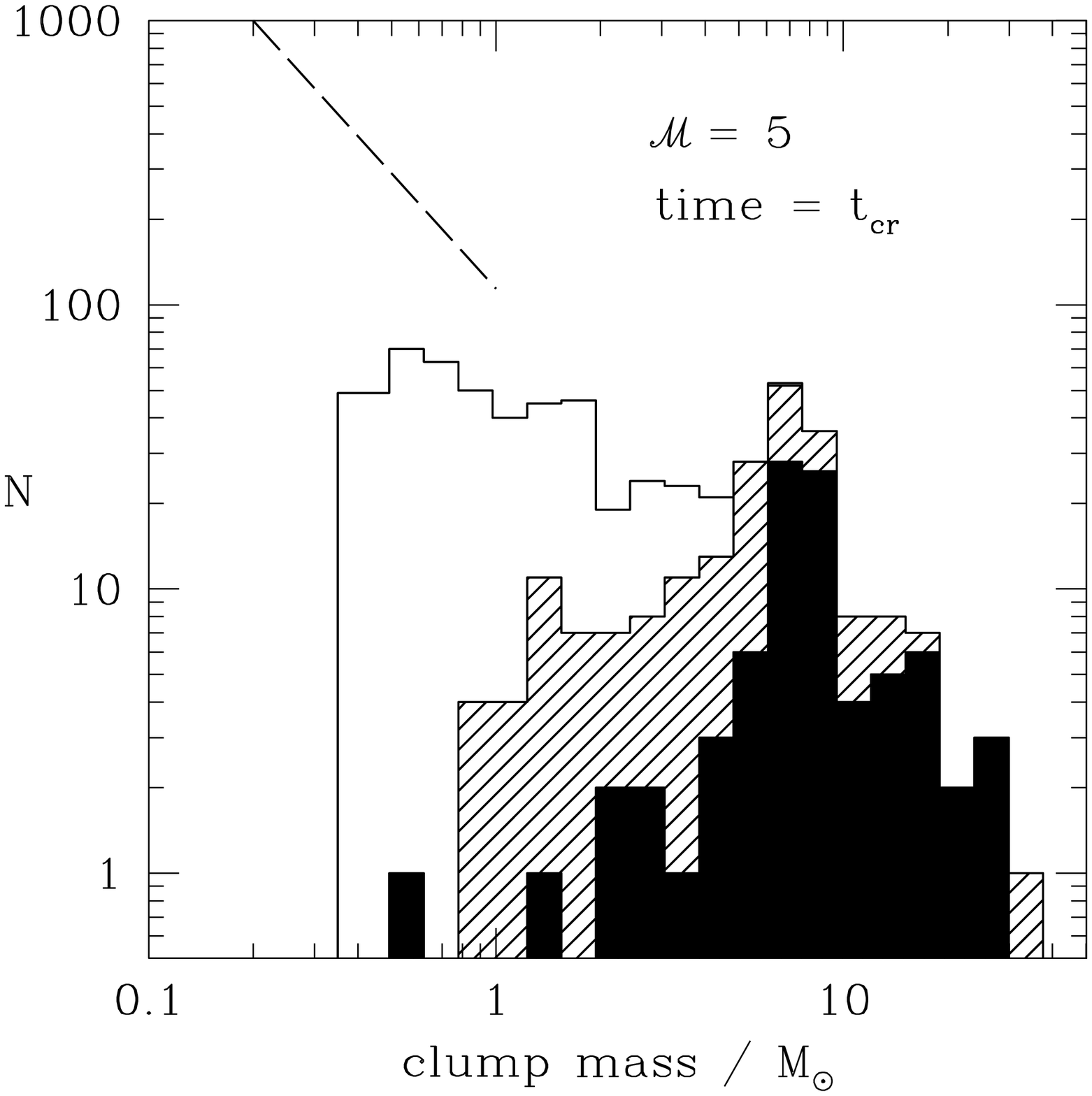}   		}
\centerline{	\includegraphics[width=2.8in,height=2.8in]{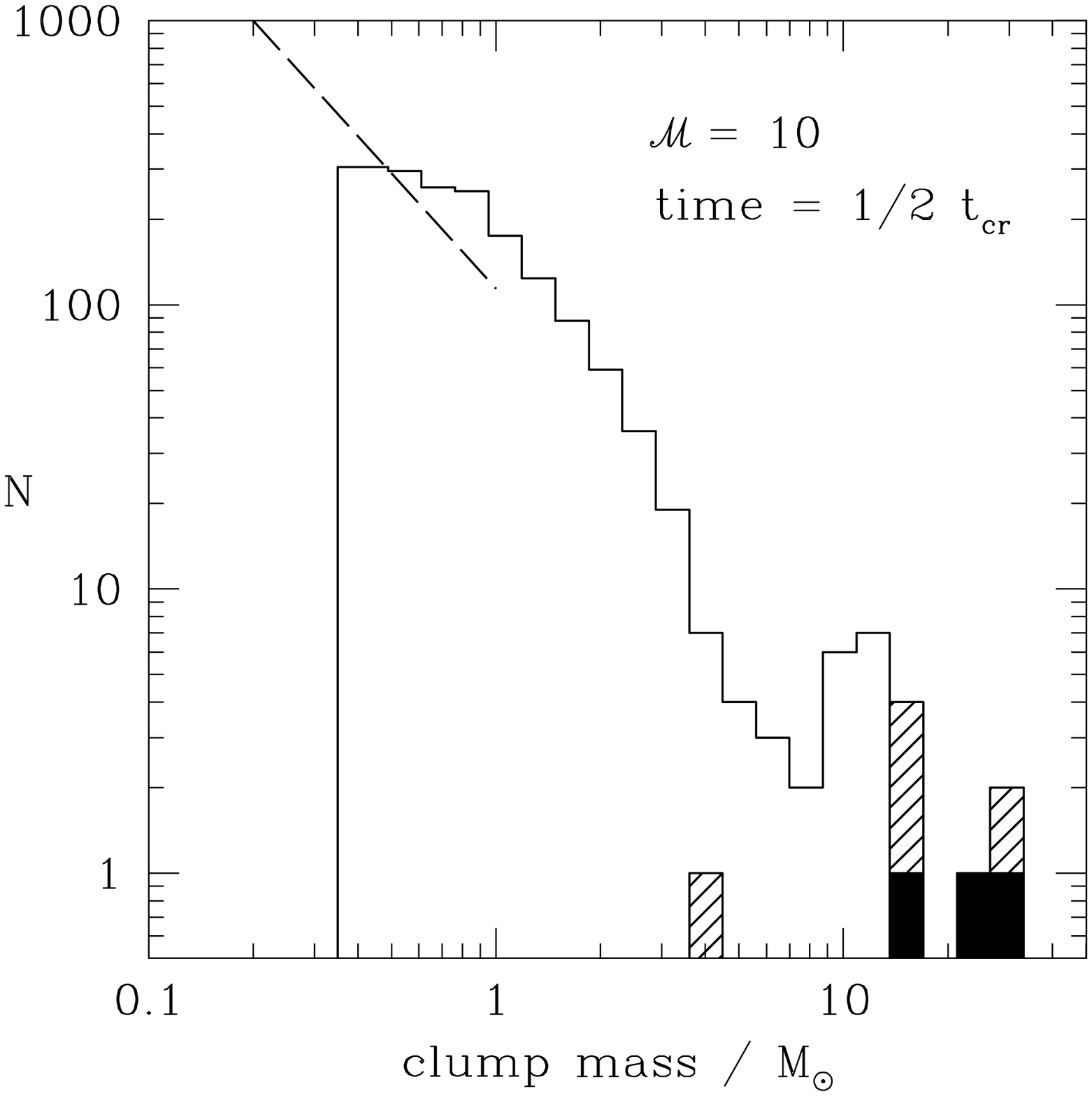} 
			\includegraphics[width=2.8in,height=2.8in]{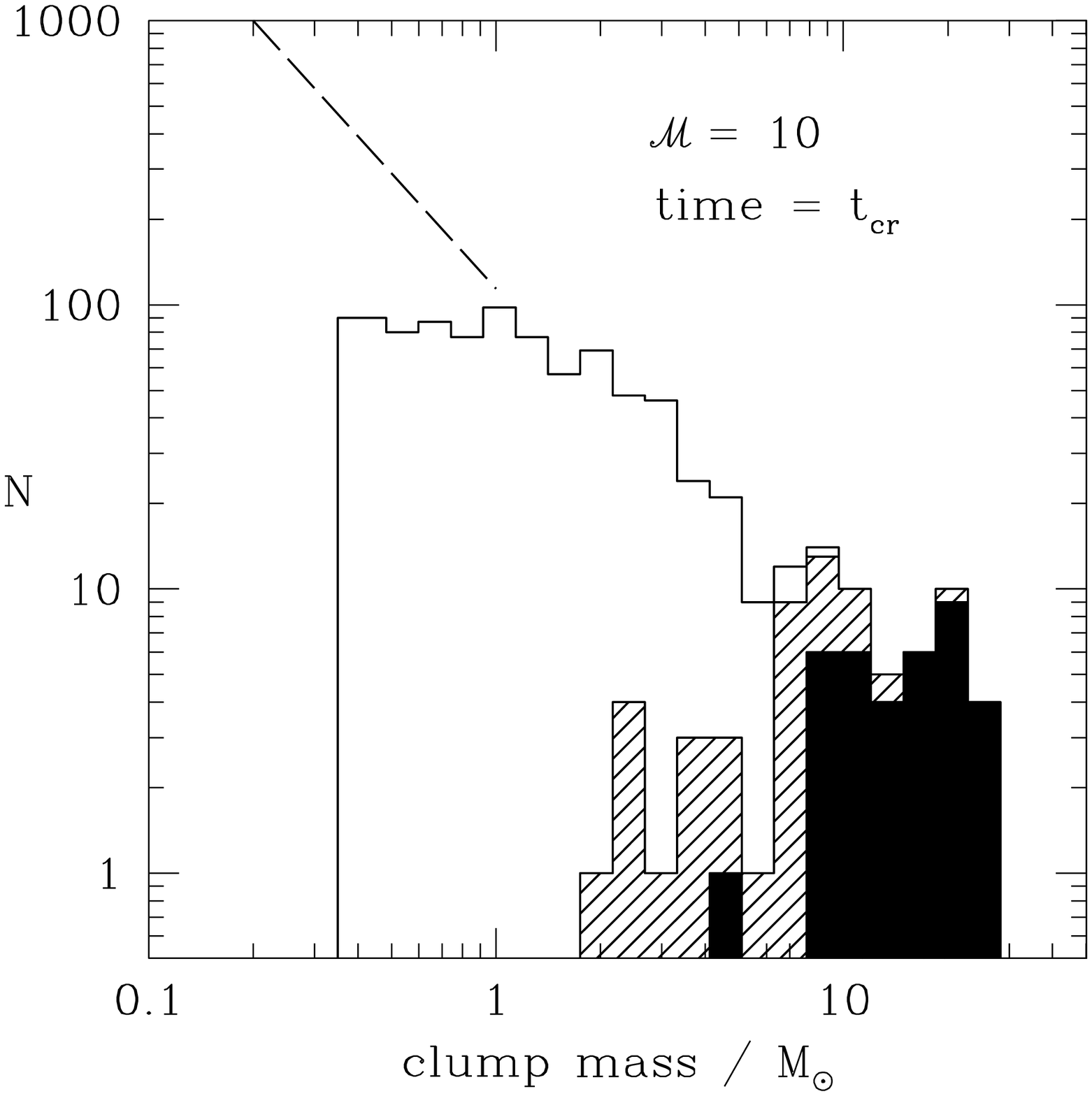} 		} 
\centerline{ 	\includegraphics[width=2.8in,height=2.8in]{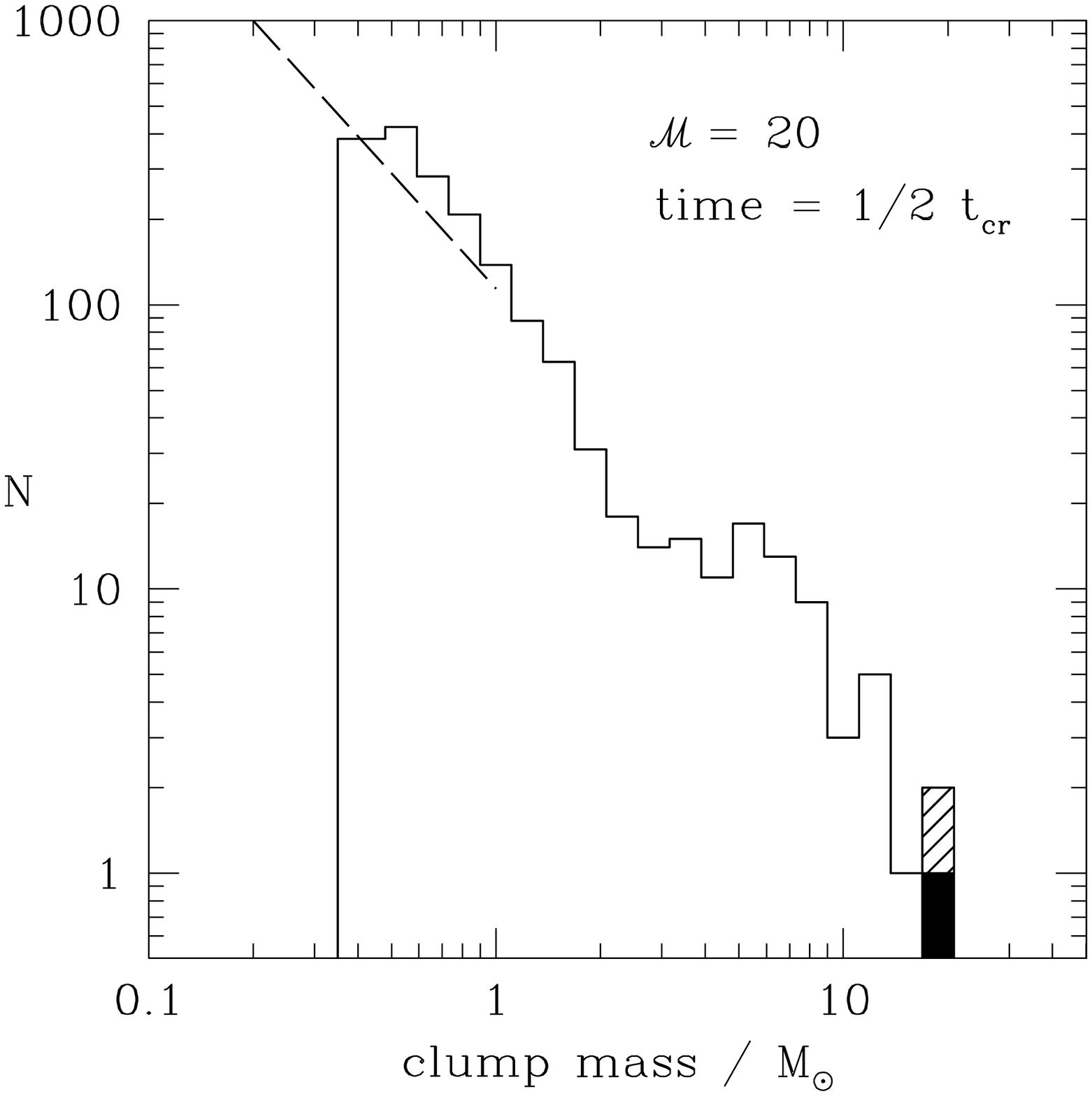} 
			\includegraphics[width=2.8in,height=2.8in]{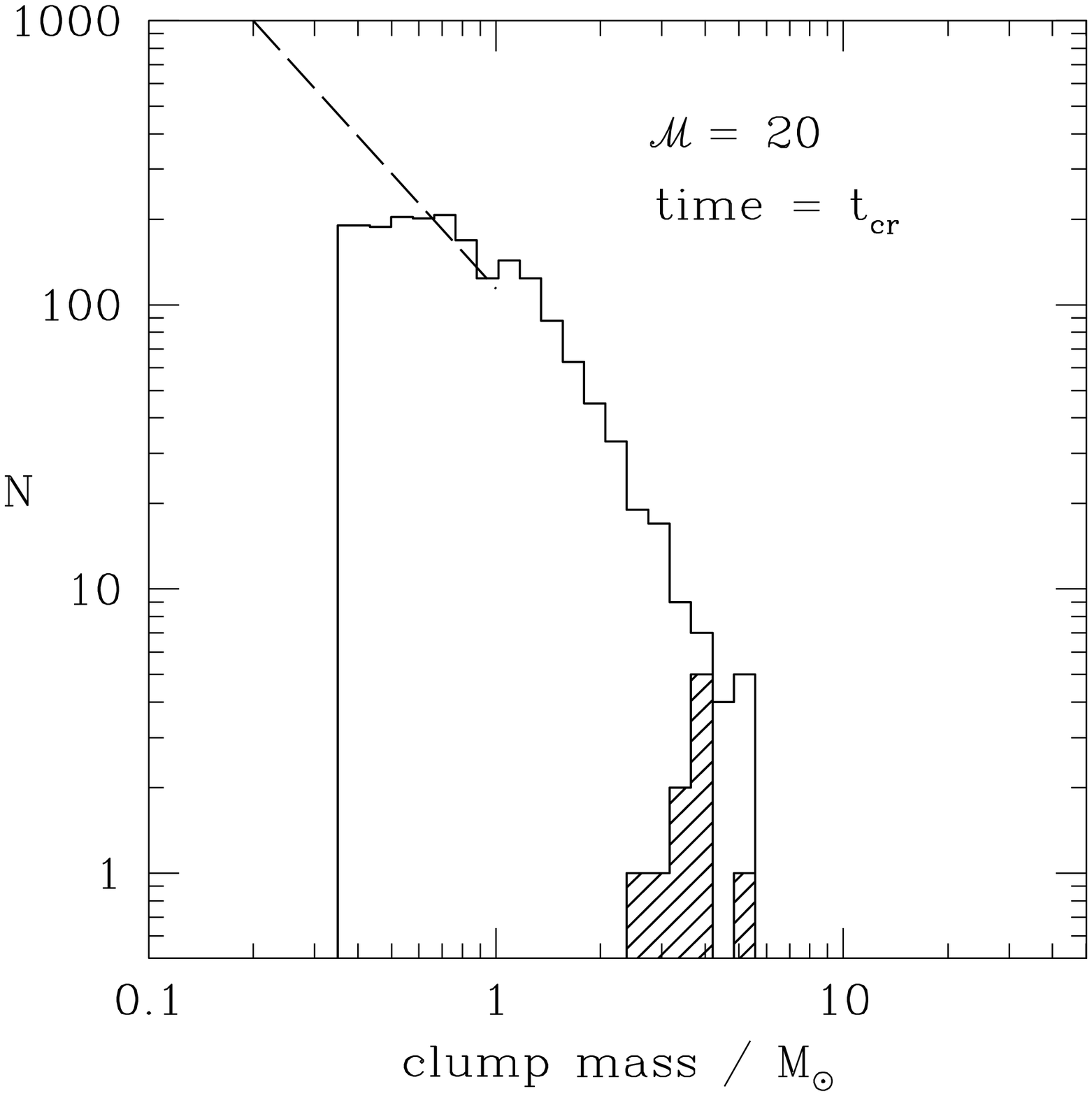} 		}
\caption{\label{gravspectra} In this figure we show the clump mass spectra for the 40\% filling
factor flows but this time, the self gravity of the gas is included in the simulations. The un-shaded 
region denotes the full clump distribution. The hatched region is the sub set of clumps for which
$|E_{grav}| \ga E_{thermal}$ and the black regions denote the clumps for which $|E_{grav}| \ga (E_{thermal} + E_{kinetic})$. Again, the clump spectra are plotted at 0.5\tcr~and \tcr. The long 
dashed line shows a Salpeter slope of $\gamma = 2.35$ (so $-1.35$ on our log-log plot here). }
\end{figure*}

\section{The influence of gravity}
\label{gravsection}

In the previous section we showed that clumpy colliding flows can produce a
region with a clump mass function that, in the absence of self-gravity, resembles 
the stellar IMF. We find that
shredding dominates over coagulation in all these simulations, however in our 
lower Mach number runs (\Mach = 5) we see that some steady coagulation does occur
and creates a number of higher mass objects. In the higher Mach number runs
(\Mach = 10 \& 20), higher mass clumps are also present at the point in the
simulations where the cloud is at its most dense ($t = 0.5$\tcr). These are only transient
features, however, and most are gone by $t = $\tcr.

In this section we examine the effects of self-gravity in some of these runs. We 
have only selected 3 simulations in this section, to be
re-run with the self-gravity switched on, since the computational expense of this
type of calculation is much greater. We have chosen to re-run the 40\% filling
factor simulations, with Mach numbers 5, 10 and 20. We have chosen these
simulations since at Mach 10 and Mach 20, their non-self-gravitating versions
give mass spectra that are remarkably similar in slope, at $t = 0.5$\tcr, to the stellar 
IMF.

The three 40\% filling factor simulations were re-run with the self-gravity switched on.
Just like the original versions, they were allowed to run for one crossing time.
Figure \ref{gravpics} shows column density images from the {\it{end}} of each of the three 
simulations. The Mach 5 and Mach 10 flows both resulted in star formation during the
crossing time but no star formation occurred in the Mach 20 flow. In the simulation
with the Mach 5 flow, the star formation occurred at $t = 0.45$ \tcr~( or 1.8 $\times 10^{6}$ 
years), and in the Mach 10 simulation at $t = 0.63$ \tcr~(or  $1.2 \times 10^{6}$ years). 

Figure \ref{gravspectra} shows the clump mass spectra for these new simulations.
At $t = 0.5$\tcr~the spectra look very similar to previous simulations where 
self-gravity was excluded. The Mach 10 and 20 flow simulations thus have the same 
Salpeter type slope, and now so too does the Mach 5 flow simulation. 
At $t = $\tcr~the Mach 5 and 10 simulations have an overall flatter 
distribution than before, having more high mass clumps,  and fewer low mass clumps.
The Mach 20 simulation has essentially the same distribution.

The inclusion of self-gravity to the calculations therefore did not manage to allow
the Mach 20 simulation to preserve the more massive clumps that are formed during
the flow. However, its inclusion also did not significantly alter our previous result, that
during the densest phase of the flows, that is time $t = 0.5 $\tcr, the mass spectra of the
region has a Salpeter type slope.

The fact that the clump mass spectra for the Mach 5 and 10 flows is flatter at $t = 
$\tcr than in the non-self-gravitating case is simply a result of the flows collapsing 
under self gravity.
This collapse can be seen clearly in the column density images in figure \ref{gravpics}.
Instead of having a 1-D compression of the gas along the $x$ axis, as occurred in the 
non-self-gravitating cases, the simulations with self-gravity experience a a 3-D 
compression. This results in more coagulation of smaller clumps into larger structures.
The Mach 20 simulation is less affected by this gravitational collapse, since the flow
crossing time is much shorter than the gravitational free-fall time of the flows: it simply
has less time to feel the effects of the self gravity.

It is interesting that the Mach 5 clump distribution is also more Salpeter like with 
the inclusion of gravity. If we look back at the 10\% filling factor runs in figure
\ref{clumpspectra}, we see that they are always steeper than those for the higher filling 
factor runs. This is probably a result of the higher initial clump density in the low
filling factor runs (see section \ref{setup} for details of how the runs are set up). At 
these higher densities, the shredding is likely to be more efficient, due to the 
increased the ram pressure.
Similarly, in the 40\% filling factor run, with the self gravity included, the collapse of 
the flow under the gravity will create denser clumps, perhaps aiding the initial 
shredding, before the flow collapse can result in coagulation. 

Figure \ref{gravspectra} also plots the distribution of clumps for which the gravitational
energy is greater than the thermal energy, which we call {\it{thermally bound}}, and the
clump distribution for which the gravitational energy is greater than both the supporting
kinetic and thermal energies. We will refer to these latter clumps as {\it{totally bound}}. 
We must remember here that none of the structures initially have any internal kinetic 
energy. The kinetic energy they now possess has come from the shock formation process.
The thermally bound population generally spans a wider range in clump mass than the
totally bound clumps. It is only the more massive clumps in the distribution that manage
to have little enough kinetic energy to become totally bound. The star formation that
occurs in these simulations is therefore confined to the most massive of the clumps. 
Note that these star forming clumps do not constitute a representative subset of the 
main clump distribution. In fact, a similar result was seen in simulations of both driven 
\citep{Klessen2001} and freely decaying turbulence \citep{ClarkBonnell2005}, in that
in both cases the star formation was confined to largest clumps.

The fact that the bound clumps form at or around 20 \solmas shows that the shock 
clump-formation process does little to alter the energy state of the original clump gas. 
The clumps originally had a Jeans number of 0.5 \solmasp, which with an original clump
mass of 6 \solmasp, gives the original Jeans mass in the clump gas to be 17 \solmasp.
Thus only clumps that have a mass similar to the original Jeans mass become bound.
This is consistent with the findings of \citep{ClarkBonnell2005}, who showed the same
result for freely decaying turbulence.

As already mentioned, the Mach 20 flow simulation failed to produce any stars in a time
period of one crossing time. Interestingly, it did have one bound clump at $t = 0.5 $\tcr,
which (again) was among the most massive of the clumps in the simulation, but this
clump was lost as the flow evolved. At the point at which the simulation was stopped, 
the clump population does have quite a few lower mass members that are thermally
bound. It is possible that if the flow was left for longer that this simulation might
eventually produce some stars.

%
% Implications, comparison to observations
%

\section{Discussion}
\label{discussion}

We have proposed in this paper that the clump mass spectra that have been
observed in $\rho$ Oph, Serpens and the Orion B cloud may have arisen naturally if the
regions were formed from large scale shocks. $\rho$ Oph also displays some
evidence for a shocked formation. \citet{Motteetal1998} discuss the possibility that 
$\rho$ Oph has been triggered by a large shock from the nearby OB association 
of Sco OB2 \citep{deGeus1992}.

The morphology of $\rho$ Oph alone is suggestive of a shock formation process.
The region has a filamentary structure in the form of ridge of high
column density gas, first shown by \citet{WilkingLada1983} from $^{18}$CO 
observations. Higher density molecular tracers also revealed that this
ridge is broken up into a series of dense cores (labelled/referred to as Oph-A to
Oph-F.), \cite{LorenWootten1986, Lorenetal1990}. These cores are also clearly 
visible in the 1.3 mm continuum observations presented by \citet{Motteetal1998}.

It also appears that the region, and the star formation within, occurred fairly rapidly
in $\rho$ Oph. The young stellar objects, buried in the cores found along the 
ridge, have a relatively strong 1.3 mm continuum emission, suggesting that they
are young Class I protostars, with ages of $10^{5}$ years. This is an order of 
magnitude less than the crossing time along the ridge. suggesting that the star 
formation occurred in a fairly rapid burst. The star formation that occurs in the 
simulations we present in this
paper also occurs fairly rapidly, on a timescale less that the flow crossing time.
Both the observations of $\rho$ Oph and the simulations
presented here are therefore consistent with suggestion by \citet{Elmegreen2000},
that star formation occurs on a timescale comparable to the crossing time in the
region.

Although the triggering for the $\rho$ Oph region has so far only been attributed to the nearby
Sco OB2 association, even in this case, a scenario similar to that described here may
occur. \citet{Whitworth1979} has shown that the expansion of an ionised region into
molecular gas is preceded by a dense shocked layer in the molecular region. In this 
case however, the properties of clumps would probably be different, with those in the 
shocked layer (and in this case, the moving population) begin denser than those in
the rest of the molecular cloud. Although we have demonstrated in this paper that the 
exact form of the flows has no real influence on the resulting clump mass spectra, we
have only examined the simplest case of initially identical clumps. Further investigation
would be required to establish whether a triggering event from a nearby OB association
could produce the effect we describe in this paper.

We note here that our simulations do not contain any model for magnetic fields, and as
such, we can say nothing about such fields would affect the clump mass distribution.
\citet{CliffordElmegreen1983} have shown that magnetic linking between clumps can
increase the chance of clump-clump interactions. At low Mach numbers, these magnetic
interactions may be stronger than direct clump-clump collisions, thus dominating the
processes outlined in this paper.

%
% Summary
%

\section{Conclusions}
\label{summary}

We have demonstrated that the process of colliding clumpy flows can result in a clump
mass distribution with a power law consistent with Salpeter ($\gamma = 2.35$). 
Our results highlight that the clump mass distribution observed in the $\rho$ Oph 
\citep{Motteetal1998, Johnstoneetal2000}, Serpens \citep{TestiSargent1998} and the Orion B 
\citep{Johnstoneetal2001} molecular clouds, need have nothing to do with the on going
star formation. Instead, clumpy structure may arise naturally if these clouds were formed 
from large scale clumpy shocks.

Our results are largely indifferent to our choice of flow Mach numbers and filling factors,
or even whether gravity is included in the model, or not. This suggests that the structure
and kinematics of the gas which forms star forming regions, is not important in forming
the observed clump/core mass distribution.

The mass spectra in our simulations are not however constant in time. Only at the 
densest point of the shock process (i.e. when the full length of the flows have entered
the shock region) do all the mass spectra yield the Salpeter type slope. This is however,
the point at which our self-gravitating simulations show us that star formation sets in, and
the spectra exist for the embedded phase: the time period in which we are likely to try
and carry out  observations of clump/core populations.

\section{Acknowledgements}

Paul Clark would like to acknowledge a UKAFF Fellowship. We would also like to
thank Jim E. Pringle, and the referee, Bruce G. Elmegreen, for useful comments and 
suggestions which greatly improved the paper.

%
% bibliography using bibtex and natbib...
%
 
\bibliographystyle{mn2e}
\bibliography{/Users/paul/Documents/bibfile/pccbib}

\end{document}